\documentclass[10pt, aps, prd, preprintnumbers, superscriptaddress, nofootinbib, amsmath,amssymb]{revtex4-2}

\usepackage{style}
\usepackage{aas_macros}

\begin{document}

\preprint{DESY-25-122}

\title{Prospects and Limitations of PTAs Anisotropy Searches -- The Frequentist Case}

\author{Thomas Konstandin\,\orcidlink{0000-0002-2492-7930}}
 \email{thomas.konstandin@desy.de}
\affiliation{Deutsches Elektronen-Synchrotron DESY, Notkestr. 85, 22607 Hamburg, Germany}

\author{Anna-Malin Lemke\,\orcidlink{0009-0005-3568-3336}}
 \email{anna-malin.lemke@desy.de}
\affiliation{II. Institute of Theoretical Physics, Universität Hamburg, Luruper Chaussee 149, 22761, Hamburg, Germany}
\affiliation{Deutsches Elektronen-Synchrotron DESY, Notkestr. 85, 22607 Hamburg, Germany}

\author{Andrea Mitridate\,\orcidlink{0000-0003-2898-5844}}
 \email{andrea.mitridate@desy.de}
\affiliation{Deutsches Elektronen-Synchrotron DESY, Notkestr. 85, 22607 Hamburg, Germany}

\author{Enrico Perboni\,\orcidlink{0009-0002-5799-7625}}
 \email{enrico.perboni@desy.de}
\affiliation{Deutsches Elektronen-Synchrotron DESY, Notkestr. 85, 22607 Hamburg, Germany}

\begin{abstract}
Recent findings from several Pulsar Timing Array (PTA) collaborations point to the existence of a Gravitational Wave Background (GWB) at nanohertz frequencies. A key next step towards characterizing this signal and identifying its origin is to map the sky distribution of its power. Several strategies have been proposed to reconstruct this distribution using PTA data. In this work, we compare these different strategies to determine which one is best suited to detect GWB anisotropies of different topologies. We find that, for both localized and large-scale anisotropies, reconstruction methods based on pixel and radiometer maps are the most promising. However, in both scenarios, even the optimistically large anisotropic signals discussed in this work remain challenging to detect with near-future PTA sensitivities. For example, we find that for a GWB hotspot contributing to $80\%$ of the GWB power in the second frequency bin, detection probabilities reach at most $\mathcal{O}(10\%)$ for a PTA with noise properties comparable with the ones of the upcoming IPTA third data release. Finally, we consider the fundamental limitations that cosmic variance poses to these kinds of searches by deriving the smallest deviations from isotropy that could be detected by an idealized PTA with no experimental or pulsar noise.
\end{abstract}

\maketitle
\tableofcontents
\clearpage
\section{Introduction}
All regional Pulsar Timing Array (PTA) collaborations have reported evidence pointing to the existence of a Gravitational Wave Background (GWB) at nanohertz frequencies~\cite{NANOGrav:2023gor,EPTA:2023fyk,Reardon:2023gzh,Xu:2023wog,Miles:2024seg}.
As more data is collected, it is important to develop and test the tools that will be needed to characterize this GWB.
To date, the only property of the GWB that has been measured is its power spectrum. However, as more data comes online, a more granular characterization will be possible. 
For example, future data sets will allow us to test with growing precision whether the gravitational wave background is statistically isotropic. Detecting deviations from isotropy would be a crucial next step in the identification of the GWB source. Indeed, the most plausible source for the GWB -- a population of supermassive black hole binaries (SMBHB)-- is expected to produce localized anisotropies corresponding to the location of bright binaries (see, for example, Ref.~\cite{Taylor:2013esa, Gardiner:2023zzr, Lemke:2024cdu}). On the other hand, most cosmological sources would produce a GWB whose anisotropies are well below the sensitivity of current and future PTAs. Therefore, detecting deviations from an isotropic distribution would provide strong evidence in favor of an astrophysical origin of the signal. 

PTA searches for anisotropies aim to detect deviations from the expected correlations between the timing signals of different pulsars. For an isotropic gravitational wave background, these correlations are a simple function of the pulsars’ angular separations, known as the Hellings-Downs (HD) correlation function. On the other hand, anisotropies in the GWB would induce deviations from these HD correlations in a way that can be used to search for them. These searches have been carried out using both Bayesian and frequentist techniques. In the former approach (see, for example, Refs.~\cite{Taylor:2015udp, NANOGrav:2023tcn}), Bayesian techniques are used to analyze the measured timing residuals and reconstruct the GWB sky map. In the frequentist approach, the information encoded in the timing residuals is first condensed into a set of estimators for the pulsar cross-correlations, which are then used to test the isotropic assumption (see, for example, Ref.~\cite{Pol:2022sjn}). In this work, we focus on frequentist searches, leaving the analysis of Bayesian search strategies to future work. 

As we will discuss in this work, any frequentist anisotropy search consists of two main steps. First, a GWB sky map is reconstructed from PTA data. Second, this map is used to define a detection statistic that quantifies the statistical significance of any deviation from isotropy. In each of these steps, several approaches can be taken: the map reconstruction can be performed under various parameterizations, and various detection statistics can be employed to assess the presence of anisotropy. The goal of this work is to answer these questions:
\begin{enumerate}
    \item Which combination of map parametrization and detection statistic is best suited to detect GWB anisotropies with different topologies? 
    \item What are the detection prospects in current and future PTA data sets?
    \item What is the fundamental limit introduced by cosmic variance in our ability to detect GWB anisotropies? 
\end{enumerate}

Compared to previous works that benchmarked different anisotropy search strategies and derived detection forecasts~\cite{Pol:2022sjn, Lemke:2024cdu, Depta:2024ykq}, our work is the first one to (1) Properly account for the impact of cosmic variance on anisotropy detection prospects \cite{Konstandin:2024fyo} (2) Perform a systematic comparison of different combination of map reconstruction methods and detection statistics (3) Forecast the detectability of anisotropies with different topologies (4) Include pair-covariance in the map-reconstruction procedure~\cite{Gersbach:2024hcc} (5) Include in the study the recently developed frequency-resolved map reconstruction methods~\cite{Gersbach:2025mhj}.

The paper is structured as follows. In Section~\ref{sec:gwb_signal}, we review how the signal of a GWB appears in PTA data. Then, in Section~\ref{sec:timing_corr}, we review the different techniques that can be used to leverage the properties of this signal to reconstruct the GWB sky map. These different techniques are then benchmarked in Section~\ref{sec:benchmarks}. Finally, in Section~\ref{sec:cosmic_variance} we discuss the fundamental limit introduced by cosmic variance in PTA anisotropy searches. We then conclude in Section~\ref{sec:conclusions}.

\section{GWB signals in PTA data}\label{sec:gwb_signal}
Pulsar Timing Arrays monitor the radio pulses emitted by a collection of millisecond pulsars in the galactic neighborhood, typically located within a kiloparsec of Earth. By recording the times of arrival (TOAs) of these pulses with an accuracy of $\sim100$ ns, PTAs can detect the perturbations induced in these TOAs by a gravitational wave background (GWB) permeating the galaxy. In this section, we review how the features of the GWB are imprinted on the statistical properties of the TOA perturbations it generates.

Assuming that all GWB sources are far away from the Earth-pulsar systems, we can decompose the GWB metric perturbation, $h_{ij}(t,\vec{x})$, in the PTA neighborhood as a superposition of plane-waves:
\begin{equation}\label{eq:gwb_metric}
    h_{ij}(t,\vec{x})=\sum_A\int_{-\infty}^{\infty}df\int_{S^2}d\hat\Omega\; \tilde h_A(f,\hat\Omega) e^{i 2\pi f(t-\hat\Omega\cdot\vec{x})}e_{ij}^A(\hat{\Omega})\,,
\end{equation}
where $f$ is the GW frequency, $\hat\Omega$ the direction of propagation of the plane waves, $A=+,\times$ labels the two GW polarizations, $e_{ij}^A$ are the GW polarization tensors, and $\tilde h_A(f,\hat\Omega)$ are two complex functions (one for each GW polarization) satisfying $\tilde h_A^*(f,\hat\Omega)=\tilde h_A(-f,\hat\Omega)$. These metric perturbations will induce a shift, $\delta t_a(t)$, in the TOAs of the $a$-th pulsar given by (see, for example, ~\cite{Maggiore:2018sht, Taylor:2021yjx})):
\begin{equation}\label{eq:dt}
    \delta t_a(t)=\sum_A\int_{-\infty}^{\infty}df\int_{S^2}d\hat\Omega\; \tilde h_A(f,\hat\Omega) R^A_a(f,\hat\Omega) \dfrac{e^{2\pi i f t}}{2\pi i f}\,,
\end{equation}
where we have defined the response function $R_a^A(f,\hat\Omega)$ as:
\begin{equation}
    R^A_a(f,\hat\Omega)\equiv F^A_a(\hat\Omega) \left[1-e^{-2\pi i f L_a(1+\hat p_a\cdot\hat\Omega)}\right]\,,\qquad F^A_a(\hat\Omega)\equiv\frac{\hat p_a^i\hat p_a^j}{2(1+\hat\Omega\cdot \hat p_a)}e^A_{ij}(\hat\Omega)\,,
    \label{eq:ar}
\end{equation}
with $\hat p_a$ being the unit vector pointing from Earth to the $a$-th pulsar, and $L_a$ the distance from Earth to the $a$-th pulsar in our array. The first term in the square brackets of the response function corresponds to the ``Earth term'', while the exponential in the square brackets gives the ``pulsar term''.

A GWB is characterized by the fact that the functions $\tilde h_A(f,\hat \Omega)$ can be treated as random variables, drawn from some distribution that is set by the properties of the GWB source. 
Since the GWB is expected to arise as a central-limit-theorem process, it is common to model it as a Gaussian process. The one-point function of this Gaussian ensemble is usually assumed to be zero, i.e., $\langle\tilde h_A(f,\hat \Omega)\rangle=0$ where $\langle\cdot\rangle$ denotes the ensemble average. Therefore, the GWB is fully characterized by the two-point function:
\begin{equation}\label{eq:covariance}
    \langle \tilde h_A^*(f,\hat\Omega)\tilde h_{A'}(f',\hat\Omega')\rangle = \delta_{AA'}\delta(f-f')\delta(\hat{\Omega},\hat{\Omega}')H(\hat\Omega, f)\,,
\end{equation}
where $\delta_{AA'}$ arises from the assumption that the background is unpolarized,  $\delta(f-f')$ implies that the background is stationary in time, and $\delta(\hat{\Omega}, \hat{\Omega}')$ implies that the background is homogeneous. The GWB power spectrum, $H(\hat{\Omega},f)$, can be factorized as $H(\hat{\Omega},f)=H(f)P(\hat{\Omega},f)$, where the function $H(f)$ describes the spectral content of the GWB, and $P(\hat{\Omega},f)$ describes the distribution of the GWB power in the sky and is normalized such that $\int d\hat\Omega \, P(\hat\Omega,f)=4\pi$.

From Eq.~\eqref{eq:dt} and Eq.~\eqref{eq:covariance}, it then follows that the timing residuals produced by a GWB follow a Gaussian distribution with zero mean. The two-point function of this Gaussian distribution is given by:
\begin{equation}\label{eq:res_corrs}
    \langle \delta t_a(t_i)\delta t_b(t_j)\rangle = \int_{-\infty}^\infty df\; \rho_{ab}(f)\Phi(f)e^{2\pi i f (t_j-t_i)}\,,
\end{equation}
where we have defined the timing residuals power spectral density $\Phi(f)\equiv 2H(f)/(3\pi f^2)$, and the overlap reduction function, $\rho_{ab}$, as 
\begin{equation}\label{eq:orf}
    \rho_{ab}(f)\simeq\frac{3}{2}(1+\delta_{ab})\sum_A\int_{S^2}\frac{d\hat\Omega}{4\pi}\;F_a^A(\hat\Omega)F_b^A(\hat\Omega) P(\hat\Omega,f)\,.
\end{equation}
In deriving Eq.~\eqref{eq:res_corrs}, we have ignored the ``pulsar term" contribution to the response function for the cross-correlations. For an isotropic GWB, this is a justified assumption as long as typical pulsar distances satisfy $fL_a\gg1$; in this limit, the pulsar term contributions to the integral in Eq.~\eqref{eq:orf} average to zero. However, this assumption starts to break down when the sky is highly anisotropic. Consider, for example, the case of a GWB with a very bright hotspot. In this case, the integral in Eq.~\eqref{eq:orf} is dominated by the sky direction corresponding to the location of the hotspot, and the pulsar term no longer averages out. However, including the pulsar term in the overlap reduction function would require knowledge of the pulsar distance with a precision smaller than the GW wavelength. Given that pulsars' distances are not currently known at that level of precision, we decided not to include the pulsar term in this analysis. This is also the standard approximation made in all anisotropy searches carried out to date. Pulsar terms could be included in a Bayesian analysis, and the uncertainty on the pulsars' distances numerically marginalized over.

\section{A primer on anisotropy searches in \texorpdfstring{PTA\MakeLowercase{s}}{PTAs}}\label{sec:timing_corr}
From Eq.~\eqref{eq:res_corrs} we can see how the spectral properties of the GWB, i.e. $H(f)$, control the time-correlation of the signal, whereas the spatial distribution of the GWB power, i.e. $P(\hat\Omega,f)$, controls the correlations between the signals of different pulsars. Frequentist searches of anisotropies leverage this fact to construct an estimator of $\hat P(\hat\Omega,f)$. Specifically, they typically proceed as follows:
\begin{enumerate}
    \item The timing residuals are used to construct an estimator of the pulsar cross-correlation coefficients (Sec.~\ref{subsec:os}).
    \item From these estimated cross-correlations, they derive an estimator of the GWB sky map, $\hat P(\hat\Omega,f)$ (Sec.~\ref{subsec:sky}). 
    \item Using the reconstructed sky map, they build a detection statistic to quantify deviations from isotropy (Sec.~\ref{subsec:snr}).
\end{enumerate} 

\subsection{Estimating Cross-Correlations}\label{subsec:os}
The reconstruction of the GWB-induced cross-correlations is somewhat complicated by the fact that, in a realistic PTA experiment, sources other than the GWB will contribute to the observed timing residuals.
These noise sources are typically modeled as additional Gaussian processes contributing to the observed timing residuals, such that the two-point function of these residuals reads (see, for example, Refs.~\cite{Lentati:2012xb, NANOGrav:2023ctt}):
\begin{equation}\label{eq:timing_2p}
    \langle\bm{\delta t}_{a}\bm{\delta t}_{b}\rangle\equiv\bm{P}_{ab} = \delta_{ab}\bm{P}_{a} +(1-\delta_{ab})\bm{S}_{ab},
\end{equation}
where $\bm{\delta t}_a$ is a vector of size $N_{{\rm TOA},a}$ containing the measured timing residuals for the $a$-th pulsar, and where we have defined the auto, $\bm{P}_a$, and cross-covariance, $\bm{S}_{ab}$, matrices as:
\begin{equation}
    P_{a,ij}\equiv N_{a,ij}+F_{a,ik}(\Phi+\varphi_a)_{kk'}F_{a,jk'}\,,\qquad\qquad
    S_{ab,ij}=\rho_{ab,k}\, F_{a,ik}\,\Phi_{kk'} F_{b,jk'}\,.
\end{equation}
Here, the $i$ and $j$ indices run over the TOAs, $k$ runs over a set of discrete frequencies $f_k=k/T_{\rm obs}$, and we have defined $\rho_{ab,k}\equiv(\rho_{ab}(f_1), \rho_{ab}(f_1), \rho_{ab}(f_2), \rho_{ab}(f_2), \ldots)$. 
The white noise contributions to the auto-covariance matrix are encoded in a $N_{{\rm TOA},a}\times N_{{\rm TOA},a}$ matrix, $\bm{N}_a$. The contributions from the GWB and pulsar intrinsic red noise have been written in terms of a discrete Fourier basis, $\bm{F}_a$, and diagonal matrices, $\bm{\Phi}$ and $\bm{\varphi}_a$, whose diagonal entries are given by the timing residuals power spectral density induced by these two processes. For the case of the GWB, we have $\bm{\Phi}={\rm diag}(\Phi(f_1), \Phi(f_1), \Phi(f_2), \Phi(f_2), \ldots)$, where $\Phi(f)$ is related to the GWB power spectrum according to the relation given below Eq.~\eqref{eq:res_corrs}). The Fourier design matrix for the $a$-th pulsar, $\bm{F}_a$, is a $N_{{\rm TOA},a}\times 2N_f$ matrix of the form
\begin{equation}
    \bm{F_a} = \left[\begin{array}{ccccc}
         \sin(2\pi\,t_{a,1} f_1)&\cos(2\pi\,t_{a,1}f_1)&\ldots & \sin(2\pi\,t_{a,1}f_N)&\cos(2\pi\,t_{a,1}f_N)  \\
         \vdots & & \ddots & & \vdots\\
         \sin(2\pi\,t_{a,N_{\rm TOA}} f_1)&\cos(2\pi\,t_{a,N_{\rm TOA}}f_1)&\ldots & \sin(2\pi\,t_{a,N_{\rm TOA}}f_N)&\cos(2\pi\,t_{a,N_{\rm TOA}}f_N)
    \end{array}\right]
\label{eq:fourier_design}
\end{equation}
where $t_{a,i}$ are the TOAs for the $a$-th pulsar.

\emph{\textbf{Broadband estimator}} -- A common assumption in PTA anisotropy searches (see, for example, Refs.~\cite{ Mingarelli:2013dsa, Taylor:2015udp, NANOGrav:2023tcn}) is to model the sky-distribution of the GWB power as being frequency-independent, i.e., $P(\hat\Omega,f)=P(\hat\Omega)$. In this case, the cross-correlation coefficients also become frequency-independent, i.e., $\rho_{ab}(f)=\rho_{ab}$, and their optimal estimator is given by 
~\cite{NANOGrav:2023icp, Gersbach:2024hcc}:
\begin{equation}\label{eq:os_broad}
    \hat\rho_{ab}= \bm{\delta t}_a^T\cdot \bm{w}_{ab}\cdot \bm{\delta t}_b
    \qquad\qquad {\rm with} \qquad\qquad 
    \bm{w}_{ab}=\frac{\bm{P}_a^{-1}\cdot\skew{5}\tilde{\bm{S}}_{ab}\cdot\bm{P}_b^{-1}}{{\rm tr}\left[ \bm{P}_a^{-1}\cdot\skew{5}\tilde{\bm{S}}_{ab}\cdot\bm{P}_b^{-1}\cdot\skew{5}\tilde{\bm S}_{ab}\right]}\,,
\end{equation}
where we have defined the scaled cross-covariance matrix as $\skew{5}\tilde{\bm{S}}_{ab}\equiv \bm{F}_a\hat{\bm{\Phi}}\bm{F}_b$, with $\hat{\bm{\Phi}}\equiv \bm{\Phi}/A_{\rm gw}^2 $ encoding the spectral template of the GWB (see below for a discussion of the spectral templates considered in this work). 
The cross-correlation estimators derived in this way will be correlated, with a covariance matrix given by:
\begin{equation}\label{eq:rho_cov}
    \bm{\Sigma}_{ab,cd}\equiv\langle \hat\rho_{ab} \hat\rho_{cd}\rangle-\langle\hat\rho_{ab}\rangle\langle\hat\rho_{cd}\rangle={\rm tr}\left[ \bm{w}_{ab} \bm{P}_{ac}\bm{w}_{cd}\bm{P}_{db}\right] + {\rm tr}\left[ \bm{w}_{ab} \bm{P}_{ad}\bm{w}_{dc}\bm{P}_{cb}\right]\,.
\end{equation}
\emph{\textbf{Per-frequency estimator}} -- The broadband estimator defined in Eq.~\eqref{eq:os_broad}, while commonly adopted, is not the most appropriate for GWB sourced by SMBHB, as the shot noise in the power distribution produced by loud binaries will generally produce frequency-dependent anisotropies. Because of this, recent works~\cite{Gersbach:2024hcc, Grunthal:2024sor, Gersbach:2025mhj} have developed a formalism to reconstruct frequency-resolved cross-correlations, and from those frequency-resolved GWB sky maps. The optimal estimator of the frequency-dependent cross-correlation coefficients is given by 
~\cite{Gersbach:2024hcc, Gersbach:2025mhj}: 
\begin{equation}\label{eq:os_narrow}
    \hat\rho_{ab,k}= \bm{\delta t}_a^T\cdot \bm{w}_{ab,k}\cdot \bm{\delta t}_b
    \qquad\qquad {\rm with} \qquad\qquad 
    \bm{w}_{ab,k}=\frac{\bm{P}_a^{-1}\cdot\skew{5}\tilde{\bm{S}}_{ab,k}\cdot\bm{P}_b^{-1}}{{\rm tr}\left[ \bm{P}_a^{-1}\cdot\skew{5}\tilde{\bm{S}}_{ab,k}\cdot\bm{P}_b^{-1}\cdot\skew{5}\tilde{\bm S}'_{ab,k}\right]}\,,
\end{equation}
where we have defined $\skew{5}\tilde{\bm{S}}'_{ab,k}\equiv \bm{F}_a{\bm{\Phi}}\bm{F}_b^T/\Phi(f_k)$ and $\skew{5}\tilde{\bm{S}}_{ab,k}\equiv \bm{F}_a\tilde{\bm{\phi}}_k\bm{F}_b^T$, with $\tilde{\bm \phi}_k$ being a frequency selector of the form 
\begin{equation}
    \tilde{\bm\phi}_1={\rm diag}(1,1,0,0,\ldots,0,0), \qquad \tilde{\bm\phi}_2={\rm diag}(0,0,1,1,\ldots,0,0), \qquad \ldots \qquad \tilde{\bm\phi}_{N_f}={\rm diag}(0,0,0,0,\ldots,1,1)\,.
\end{equation}
As for the broadband case, the narrowband estimators will also be correlated with a covariance matrix $\Sigma_{ab,cd;k}$, whose full expression can be found in Ref.~\cite{Gersbach:2025mhj}.

Before proceeding, it is important to make a few comments on Eq.~\eqref{eq:os_broad} and \eqref{eq:os_narrow} and how they are implemented in realistic searches for anisotropies:
\begin{itemize}
    \item The construction of both broad and narrowband estimators, along with their associated covariance matrices, requires knowledge of several noise and GWB parameters that are not known a priori. For example, the scaled cross covariance matrix, $\skew{5}\tilde{\bm{S}}_{ab}$, requires knowledge of the GWB spectral shape. Its unscaled version, ${\bm{S}}_{ab}$, entering in the definition of the covariance matrix, is related to both the amplitude and shape of the GWB spectrum. 
    Construction of the auto-covariance matrix, $\bm{P}_a$, additionally requires knowledge of the white noise properties, i.e. the $\bm{N}$ matrix, and the power spectrum of each pulsar red-noise process, $\bm{\varphi}_a$. \\
    The white noise matrix, $\bm{N}$, is expressed in terms of EFAC, $G_\mu$, EQUAD, $Q_\mu$, and ECORR, $J_\mu$, parameters (one for each backend, here indicated by the $\mu$ index)~\cite{NANOGrav:2023ctt}:
    \begin{equation}
        N_{ij}=G_\mu^2(\sigma^2_{a,i}+Q_\mu^2)\delta_{ij}+J_\mu \delta_{e(i)e(j)}\,,
        \label{eq:wn_matrix}
    \end{equation}
    where $\sigma_i$ is the TOA uncertainty for the $i$-th TOA belonging to the backend $\mu$, and $\delta_{e(i)e(j)}$ denotes a Kronecker delta that equals 1 only when the epochs are the same for both TOAs considered and 0 otherwise. Typically, white noise parameters are fixed to their maximum-posterior values derived in single pulsar noise studies~\cite{NANOGrav:2023ctt, NANOGrav:2023gor}. In this work, when analyzing mock data, we will set white noise parameters to their injected values. \\
    The power spectra of the GWB and pulsar intrinsic red noises are usually parametrized as power laws:
    \begin{equation}
        \Phi(f)=\frac{A_{\rm gw}^2}{12\pi^2}\left(\frac{f}{{\rm yr}^{-1}}\right)^{-\gamma_{\rm gw}}\frac{{\rm yr}^3}{T_{\rm obs}}\,, \qquad\qquad \varphi_a(f)=\frac{A_a^2}{12\pi^2}\left(\frac{f}{{\rm yr}^{-1}}\right)^{-\gamma_a}\frac{{\rm yr}^3}{T_{\rm obs}}\,,
    \end{equation}
    where the amplitudes and spectral indices of these power spectra are unknown quantities to be extracted from the data. Typically, a preliminary Bayesian analysis, which models the GWB as a common uncorrelated red noise (CURN) process, is performed to extract the posterior distributions of these quantities. The natural choice would then be to set these parameters to their maximum likelihood values. However, it was shown in Ref.~\cite{Vigeland:2018ipb} that this choice can introduce a bias in the recovered estimate of the GWB parameters. Instead, in Ref.~\cite{Vigeland:2018ipb}, it was proposed to marginalize over red noise parameters by calculating the cross-correlations and their uncertainties over multiple random draws from the posterior distributions, resulting in what is referred to as the “noise marginalized optimal statistic” (NMOS).
    In this work, to make the analysis of a large number of simulated data sets possible, we will not perform this noise marginalization. Instead, we will set the red-noise and GWB parameters to their ''true values", i.e. the values of these parameters that we used to generate the mock data sets. We have checked that our results do not change significantly if noise marginalization is used instead of just fixing the noise parameters to their injected  values. This equivalence between noise marginalized and fixed noise analyses when simulating a large number of data sets has been shown in Ref.~\cite{Vigeland:2018ipb}, as long as the noise values used are derived from a common search instead of single pulsar searches.
    \item The normalization of the broadband estimator, $\hat\rho_{ab}$, is such that its expected value is given by $\langle \hat\rho_{ab}\rangle=A_{\rm gw}^2\rho_{ab}$. To normalize $\hat\rho_{ab}$ such that its expectation value is given by $\rho_{ab}$ (as defined in Eq.~\eqref{eq:orf}), an estimate of $A_{\rm gw}^2$ is performed using the unormalized $\hat\rho_{ab}$ under the assumption of isotropy~\cite{Pol:2022sjn}:
    \begin{equation}\label{eq:A_est}
        \hat A_{\rm gw}^2= \frac{\bm{\Gamma}^T\bm{\Sigma}^{-1}\hat{\bm{\rho}}}{\bm{\Gamma}^T\bm{\Sigma}^{-1}\bm{\Gamma}}.
    \end{equation}
    Here, $\bm{\Gamma}$ is an array containing the expected cross-correlation coefficients for all the pulsar pairs in the array in the isotropic case (i.e. the Hellings \& Downs correlations), $\hat{\bm{\rho}}$ is the array containing the estimated cross-correlations, and $\bm{\Sigma}$ is defined in Eq.~\eqref{eq:rho_cov}. This estimate of $A_{\rm gw}^2$ is then used to defined normalized cross-correlation coefficients, $\hat\rho_{ab}/\hat A_{\rm gw}^2$.\\
    Similarly, the per-frequency estimator, $\hat\rho_{ab,k}$, is normalized such that its expectation value is given by $\langle \hat\rho_{ab,k}\rangle = \Phi(f_k) \Gamma_{ab}$. To normalize $\hat\rho_{ab,k}$ such that its expectation value is given by $\rho_{ab}(f_k)$, an estimate of $\Phi(f_k)$ can be derived as
    \begin{equation}\label{eq:A_pf_est}
        \hat{\Phi}_k=\frac{\bm{\Gamma}^T\bm{\Sigma}_k^{-1}\hat{\bm{\rho}}_k}{\bm{\Gamma}^T\bm{\Sigma}_k^{-1}\bm{\Gamma}}\,
    \end{equation}
    where $\bm{\rho}_k$ is the array containing the estimated cross-correlations for the $k$-th frequency bin, and $\bm{\Sigma}_k$ is the covariance matrix for the cross-correlation estimators. Given this estimate, we can define a normalized version of the per-frequency cross-correlation estimators as $\hat\rho_{ab,k}/\hat\Phi_k$.\\
    In the remainder of this paper, unless otherwise specified, we will always indicate with $\hat\rho_{ab}$ and $\hat\rho_{ab,k}$  the normalized version of the cross-correlation optimal estimators. 

    \item To account for the effect of timing model fit, the $\bm{P}_a$ matrix is modified as~\cite{NANOGrav:2023icp}
    \begin{equation}
        \bm{P}_a = \bm{D}_a+\bm{F}_{a}(\bm{\Phi}+\bm{\varphi}_a)\bm{F}_{a}^T\,,
    \end{equation}
    where
    \begin{equation}
        \bm{D}_a = \bm{N}+\bm{M}_a\bm{E}\bm{M}_a^T\,.
    \end{equation}
    The design matrix, $\bm{M}$, is an $N_{\rm TOA}\times m$ matrix containing the partial derivatives of the TOAs with respect to each of the $m$ timing-ephemeris parameter contained in the timing model (evaluated at the initial best-fit value), and $\bm{E}$ is a diagonal matrix of very large values ($10^{40}$ by default). This places an improper, almost-infinite variance Gaussian prior on the timing model parameters. Upon inversion of $\bm{D}_a$, this choice marginalizes over the timing model uncertainties.
\end{itemize}

\subsection{Reconstructing Sky Maps}\label{subsec:sky}
Given a set of estimator for the pulsars cross-correlations coefficients, $\hat{\bm{\rho}}$ (or $\hat{\bm \rho}_k$ for per-frequency analyses), frequentist anisotropy searches build an estimator of the GWB sky map, $\hat {\bm P}$ (or $\hat {\bm P}_k$ in the case of frequency-resolved analyses), by maximizing the following likelihood function:
\begin{equation}\label{eq:likelihood}
    p(\hat{\bm{\rho}}|\bm{P})=\frac{\exp[-\frac{1}{2}(\hat{\bm{\rho}}-\bm{R}{\bm{P}})^T\mathbf{\Sigma}^{-1}(\hat{\bm{\rho}}-\bm{R}{\bm{P}})]}{\sqrt{\det(2\pi\mathbf{\Sigma})}}\,,
\end{equation}
where $\bm{P}$ is a vector containing the GWB power in each pixel, $\bm{\Sigma}$ is the cross-correlation covariance matrix as defined in Eq.~\eqref{eq:rho_cov}, and $\bm{R}$ is defined as:
\begin{equation}\label{eq:antenna_response}
    R_{p,ab}\equiv\frac{3}{2 N_{\rm pix}}\left[F_{a,p}^+F_{b,p}^++ F_{a,p}^\times F_{b,p}^\times\right]\,,
\end{equation}
where $p$ runs over a set of equal-area pixels of the GWB sky map, and the normalization of $\bm{R}$ is chosen such that for an isotropic sky we recover the HD correlations, i.e. $\bm{RP}=\bm{\Gamma}$ for $P_k=1$. 
Equation~\eqref{eq:likelihood} can be easily generalized to estimate frequency-resolved sky maps of the GWB, $\hat{\bm P}_k$; we just need to replace $\hat{\bm\rho}$ with $\hat{\bm\rho}_k$, and ${\bm \Sigma}$ with ${\bm \Sigma}_k$. In the remainder of this section, we will explicitly discuss the broadband case, but the formalism and results carry over straightforwardly to the frequency-resolved case with the substitutions just discussed.
Different parametrizations can be chosen for the sky map of the GWB power, $\bm{P}$. In this work, we will discuss the following parametrizations:

\begin{itemize}
    \item \noindent \textbf{\emph{Pixel basis}}~\cite{Lemke:2024cdu} \textbf{--} a natural parametrization for the GWB sky map is one where the map is divided into equal-area pixels, and each of the pixel values is viewed as an independent parameter, $P_k$. In this basis, the pixel values that maximize the likelihood in Eq.~\eqref{eq:likelihood} can be found analytically. However, analytical solutions of this type will generally contain pixels with negative power, which are, of course, unphysical. To solve this problem, in Ref.~\cite{Lemke:2024cdu}, we proposed to maximize the likelihood in Eq.~\eqref{eq:likelihood} while imposing the additional condition that each pixel value is positive. This maximization problem can no longer be solved analytically; we instead use the quadratic programming solver implemented in the \texttt{quadprog} package~\cite{goldfarbNumericallyStableDual1983}.\\
    We obtain the sky tessellation by using the \texttt{HEALPix} package~\cite{healpix}, which controls map resolution via the $N_{\rm side}$ parameter, related to the map's pixel number, $N_{\rm pix}$, as $N_{\rm pix}=12 N_{\rm side}^2$. To obtain numerically stable solutions, we limit the map resolution to $N_{\rm side}=4$ for the NANOGrav-like data set and $N_{\rm side}=8$ for the IPTA DR3-like case.

    \item \noindent\textbf{\emph{Radiometer basis}}~\cite{Ballmer:2005uw, Mitra:2007mc} \textbf{--} A commonly used parametrization for the sky map is the so-called \emph{radiometer} basis. In this basis, it is assumed that the GWB power is dominated by a single bright pixel. With this assumption, instead of reconstructing the GWB power over the entire sky, we can switch on one pixel at a time and reconstruct the power that each pixel would need in order to fit the measured cross-correlations.  These optimal pixel values can be derived analytically as 
        \begin{equation}
            \hat{\bm{P}}={\rm diag}(\bm{M})^{-1}\bm{X},
        \end{equation}
    where we have defined the Fisher information matrix, $\bm{M}=\bm{R}^{T}\bm{\Sigma}^{-1}\bm{R}$, and the ``dirty map", $\bm{X}=\bm{R}^{T}\bm{\Sigma}^{-1}\bm{\rho}$. In the radiometer basis, the inverse of the diagonal elements of the Fisher matrix can also be used as an estimate of the error associated with each of the reconstructed pixel values, i.e. $\sigma_p=({\bf M}_{pp})^{-1/2}$.
    
    The angular resolution of the reconstructed maps is limited by the number of pulsar pairs, $N_{\rm pp}$, contained in the data set according to $N_{\rm pix}\lesssim N_{\rm pp}$~\cite{Romano:2016dpx}. For the case of the NANOGrav 15-year data sets, this limits the number of pixels to $N_{\rm pix}\lesssim 2211$. Because of this, and following the same convention of Ref.~\cite{NANOGrav:2023tcn}, we set $N_{\rm side}=8$ for the reconstructed maps. For the IPTA DR3-like data sets, the number of pulsar pairs increases to $N_{\rm pp}=7140$, so that we set $N_{\rm side}=16$.
    
    \item \noindent\textbf{\emph{Square-root spherical harmonic basis}}~\cite{Pol:2022sjn} \textbf{--} In this basis, the square-root of the GWB power is parametrized in terms of the coefficients of a spherical harmonics decomposition, $a_{LM}$. Specifically, we write:
        \begin{equation}\label{eq:sqrt_basis}
            P_p = \left[P(\hat\Omega_p)^{1/2}\right]^2=\left[\sum_{L=0}^{L_{\rm max}}\sum_{M=-L}^{M=L}a_{LM}Y_{LM}(\hat\Omega_p)\right]^2\,.
        \end{equation}
    The resolution of the reconstructed maps is limited by the number of pulsars contained in the data set. Previous works~\cite{RomanoCornish2017} had claimed that maximal resolution was set by $\ell_{\rm max}\lesssim \sqrt{N_{\rm psr}}$, where $N_{\rm psr}$ is the number of pulsars in the data set. However, it was recently shown in Ref.~\cite{Domcke:2025esw} that the resolution can be pushed up to $\ell_{\rm max}\lesssim N_{\rm psr}$. However, for realistic PTA configurations, noise in the data limits the resolution to much lower values compared to these theoretical limits. Specifically, in Ref.~\cite{NANOGrav:2023tcn} it was shown that the effective resolution of the NANOGrav 15-year data set was approximately given by $\ell_{\rm max}=6$. In this work, we will use this empirical value of $\ell_{\rm max}$ for both the NANOGrav-like and IPTA DR3-like data sets. 
    
    The coefficients of the square-root decomposition, $a_{LM}$, can be related to the coefficients in the linear basis as
        \begin{equation}
            \displaystyle c_{\ell m} = \sum_{LM} \sum_{L^{\prime} M^{\prime}} a_{LM} a_{L^{\prime} M^{\prime}} \beta_{\ell m}^{LM, L^{\prime} M^{\prime}},\qquad\quad  \beta_{\ell m}^{LM, L^{\prime} M^{\prime}} = \sqrt{ \frac{(2L + 1) (2L^{\prime} + 1)}{4 \pi (2\ell + 1)}} C^{\ell m}_{LM, L^{\prime} M^{\prime}} C^{\ell 0}_{L0, L^{\prime} 0}\,,
        \end{equation}
    with $C^{\ell m}_{LM, L^{\prime} M^{\prime}}$ being Clebsch-Gordon coefficients, and $c_{\ell m}$ the coefficients of the linear map parametrization:
    \begin{equation}
        P_p=\sum_{\ell=0}^{\ell_{\rm max}}\sum_{m=-\ell}^{m=\ell}c_{\ell m}Y_{\ell m}(\hat\Omega_p)\,.
    \end{equation}
    In this basis, the maximum-likelihood solution cannot be found analytically. Instead, we derived it by using numerical optimization techniques provided by the \texttt{LMFIT} package~\cite{matt_newville_2024_12785036}, as implemented in the \texttt{MAPS} package~\cite{Pol:2022sjn}.
    \end{itemize}

\begin{figure}[t!]
\centering
    \includegraphics{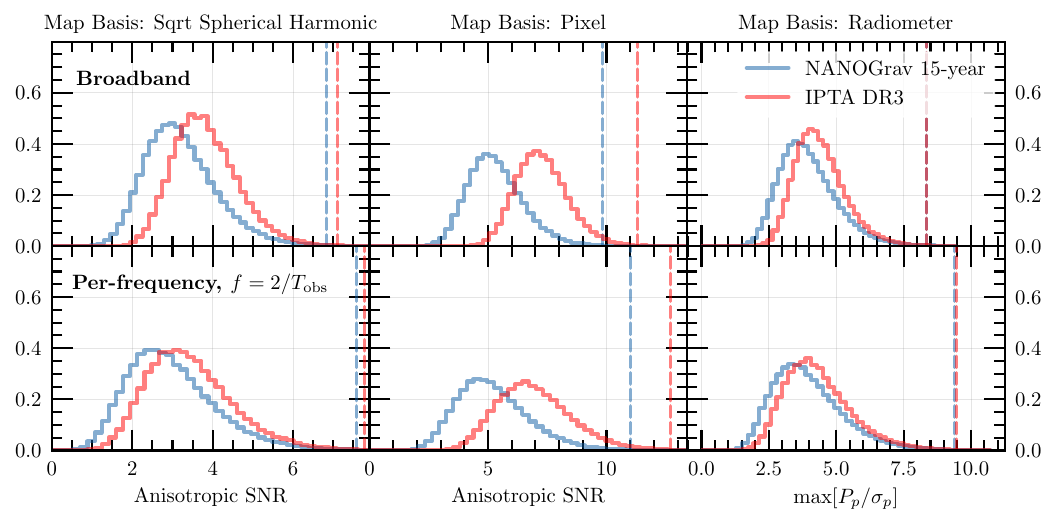}
    \caption{In the left (center) panels, we report the null distributions for the anisotropic SNR of the maps reconstructed using the square-root spherical harmonic (pixel) basis. In the right panels, we show the max radiometer SNR for maps reconstructed using the radiometer basis. The null distributions for the broadband analysis are reported in the top row, while the null distributions for the second bin of the per-frequency analysis are shown in the bottom row. In all the panels, the blue (red) histograms show the null distributions derived from a data set mimicking the noise properties of the NANOGrav 15-year (IPTA DR3) data set. The vertical dashed lines indicate the SNR value with a $p$-value=$3\times10^{-3}$ (for the per-frequency null distributions, this is a local significance, i.e. it does not take into account the look-elsewhere effect introduced by the fact that we have multiple frequency bins). In the per-frequency case, it is possible that some realizations of the GWB result in negative estimates for $A_{\rm GWB}$ in some frequency bins. Since this shows that there is no information to extract from these realizations, we include them as $\rm{SNR}=0$, resulting in small peaks separate from the bulk of the null distributions.}
    \label{fig:snr_null}
\end{figure}

\subsection{Detection Statistics}\label{subsec:snr}
Once a sky map of the GWB power has been reconstructed using one of the methods described in the previous section, the next step is to assess the statistical significance of any deviation from isotropy. This is typically done by constructing an appropriate detection statistic. A range of such statistics has been proposed in the literature. In this work, we benchmark those already adopted in anisotropy searches by regional PTA collaborations~\cite{Taylor:2015udp, NANOGrav:2023tcn, Grunthal:2024sor}:

\begin{itemize}[label=$\circ$]
    \item \emph{\textbf{Radiometer SNR}}~\cite{Pol:2022sjn,NANOGrav:2023tcn} \textbf{--} For maps reconstructed using the radiometer basis, a commonly adopted detection statistic is the value of the GWB power measured in each pixel normalized by its error, i.e., $P_p/\sigma_p$. Large reconstructed pixel values may indicate deviations from isotropy, potentially signaling the presence of a bright source emitting from that region of the sky. In this case, the detection statistic is not a scalar but an array whose size is set by the number of pixels in the reconstructed sky map. Hence, particular care will need to be taken in calibrating the detection statistic to account for look-elsewhere effects. See the discussion later in this section for details about the calibration of detection statistics.
    \item \emph{\textbf{Max Radiometer SNR}}~\cite{Grunthal:2024sor} \textbf{--} An alternative detection statistic for maps reconstructed using the radiometer basis has been proposed in the analysis of the MeerKAT 4.5-yr data set~\cite{Grunthal:2024sor}. In this work, instead of using the radiometer SNRs for all the pixels in the map, the detection statistic associated with a reconstructed sky map was defined as ${\rm SNR}={\rm max}[P_p/\sigma_p]$, where the maximum was evaluated across all the pixels of the reconstructed maps, and the error on the reconstructed pixel value, $\sigma_p$.
    \item \emph{\textbf{Anisotropic SNR}}~\cite{Pol:2022sjn} \textbf{--} A commonly adopted detection statistic is the \emph{anisotropic signal-to-noise ratio} (SNR), defined as the log-likelihood-ratio between the reconstructed GWB sky and an isotropic sky:
    \begin{equation}\label{eq:SNR}
        {\rm SNR} = \sqrt{2\ln\bigg[\frac{p(\hat{\bm{\rho}}\vert\hat{\bm{P}})}{p(\hat{\bm{\rho}}\vert \bm{P}_\text{iso})}\bigg]}\,,
    \end{equation}
    where $p$ is the likelihood function given in Eq.~\eqref{eq:likelihood}, $\hat{\bm{P}}$ is the recovered anisotropic sky map, and $\bm{P}_\text{iso}=\rm{const.}$ is a sky map with constant power across the sky. 
    \item \emph{\textbf{Spherical harmonic coefficients}}~\cite{Pol:2022sjn} \textbf{--} Finally, for maps reconstructed using the square-root spherical harmonic basis, we can use as a detection statistic the $C_\ell$ coefficients defined as:
    \begin{equation}\label{eq:Cl}
        C_\ell\equiv \frac{1}{2\ell+1}\sum_{m=-\ell}^\ell|c_{\ell m}|^2\,.
    \end{equation}
    These coefficients, commonly used in analyzing structures in the cosmic microwave background, provide a measure of the statistical fluctuations in the GWB power on angular scales $\theta\sim180^\circ/\ell$. Notice that, given our normalization of the sky maps, the monopole coefficient satisfies $C_0=4\pi$.
\end{itemize}

\begin{figure}[t!]
\centering
    \includegraphics{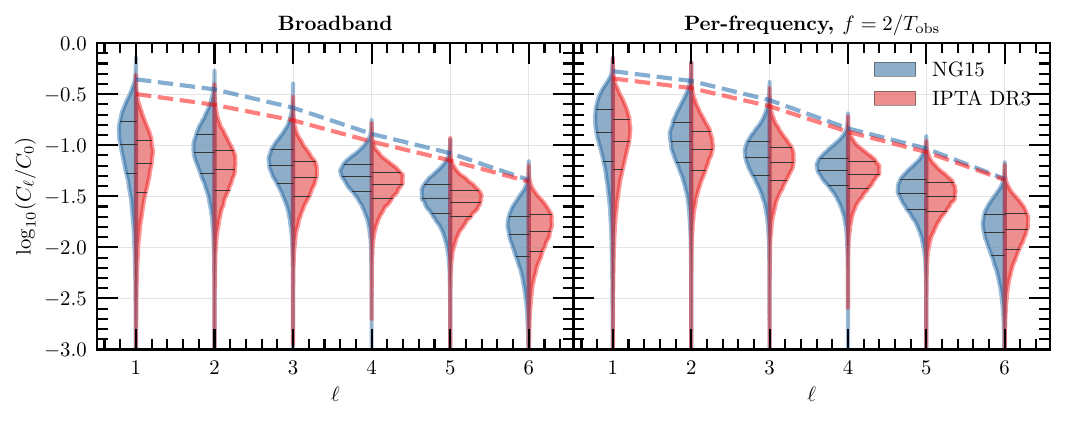}
    \caption{Null distributions for the $C_\ell$ coefficients of maps reconstructed using the square root spherical harmonics decomposition, for both the broadband (left panel) and the second bin of the per-frequency reconstruction (right panel). The blue (red) violins give the null distribution for a NANOGrav 15-year (IPTA DR3) like data set. The  dashed lines correspond to the $C_\ell$ values with a local $p$-value=$3\times10^{-3}$ (which translates to a $\sim3\sigma$ significance).}
    \label{fig:cl_null_dist}
\end{figure}

\begin{figure}[t!]
\centering
    \textbf{Broadband}\\
    \includegraphics{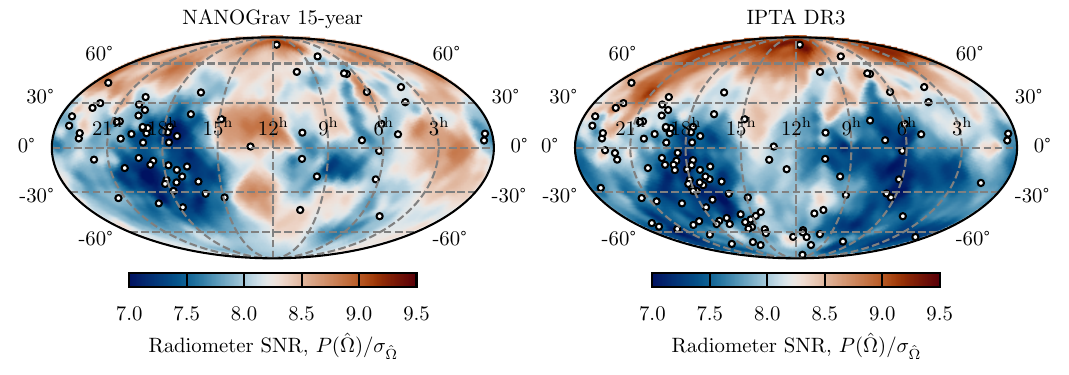}\\ 
    \textbf{Per-frequency, $f=2/T_{\rm obs}$}\\
    \includegraphics{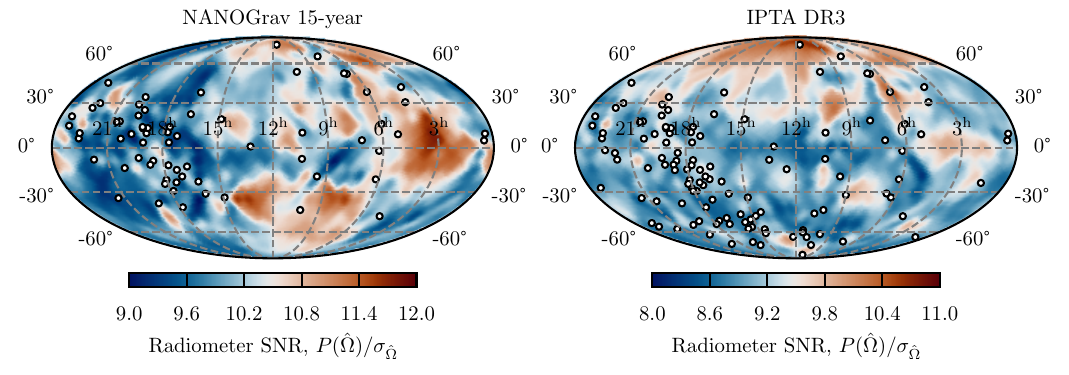}
    \caption{Radiometer SNR threshold (corresponding to a global $p$-value=$3\times10^{-3}$) for both broadband maps (upper panels) and the second frequency bin of the per-frequency reconstruction (lower panels). The left columns show the upper limits for a data set with the same noise properties as the NANOGrav 15-year data set, while the right columns show those for a data set with the noise properties of the upcoming IPTA DR3 data set. The numbers outside of the sky map represent the declination angle in degrees, while the numbers on the horizontal axis give the value of the right ascension in hours (1 hour corresponds to 15$^\circ$ of sky rotation). Before plotting, these and all other sky maps that we will show in this paper have been upscaled to a resolution of $N_{\rm side}=64$ and smoothed with a Gaussian symmetric beam with a full width at half maximum of $5^\circ$.}
    \label{fig:map_upper}
\end{figure}

However, by itself, the measured value of a detection statistic does not answer the question: is the reconstructed GWB sky incompatible with the isotropic assumption? To answer this question, we need to calibrate the detection statistics, i.e. derive (or estimate) the probability distribution of the test statistic under the null hypothesis (i.e. isotropic GWB), also known as the \emph{null distribution}. Once this distribution is known, we can obtain the $p$-value of the measured detection statistic, and quantitatively assess the statistical significance of any deviation from isotropy present in the reconstructed GWB sky map. 

The general procedure used to estimate null distributions follows three steps: (1) Several mock realizations of the PTA data are generated under the assumption of isotropy (2) Each of these realizations is run through the full anisotropy search pipeline to derive a value of the detection statistic (3) The collection of the detection statistic values derived in this way provides an estimate of the null distribution. However, different approaches have been taken in the literature when it comes to generating the mock data sets used to derive null distributions. In the latest NANOGrav analysis~\cite{NANOGrav:2023tcn}, mock data realizations were generated at the level of the pulsar cross-correlations. These were treated as independent random variables and generated by drawing from normal distributions centered around the HD values and with standard deviations given by the diagonal elements of Eq.~\eqref{eq:rho_cov} in the weak signal limit (i.e., assuming that $\bm{P_a}\gg \bm{S}_{ab}$). However, as we pointed out in Ref.~\cite{Konstandin:2024fyo}, this procedure fails to account for correlations between the cross-correlation estimators, as well as deviations from Gaussianity. Therefore, it leads to null distributions that severely overestimate $p$-values. In the latest MeerKAT search~\cite{Grunthal:2024sor}, these mock data samples were generated at the dirty map level, using the pair-covariance matrices from per-frequency optimal statistic reconstructions. While this procedure takes into account the correlations between the different estimators, it still does not account for any non-Gaussianities in the distribution of the cross-correlation estimators.

In this work, we generate mock data at the level of the timing residuals. Specifically, we use the software package \texttt{pta-replicator}~\cite{pta_replicator} (or a modified version optimized for faster data generation) to create sets of simulated pulsars' TOAs in which we inject the signal produced by an isotropic GWB, on top of Gaussian white noise and intrinsic pulsar red noise (see App.~\ref{app:mock_data} for details on the mock data generation procedure). For each of the $\sim10^6$ timing data realizations in this catalog, we then use the \texttt{DEFIANT}~\cite{Gersbach:2024hcc} implementation of the optimal estimator discussed in Sec.~\ref{subsec:os} to estimate the cross-correlation coefficients, $\hat{\bm{\rho}}$ (or their per-frequency analogue), and their associated pair-covariance matrix, $\bm{\Sigma}$.\footnote{The exact number of realizations used to generate the null distributions and calibrate the detection statistic changes from detection statistic to detection statistic. For the radiometer basis, where look-elsewhere effects are particularly important and we have to sample the low probability tail of each pixel null distribution, we use $\sim10^6$ samples. For all the other detection statistics, where look-elsewhere effects are less severe, we use $\sim10^5$ samples.} Finally, these estimated cross-correlations --and their associated pair-covariance matrices-- are used to reconstruct the GWB sky map and compute detection statistic values. We repeat this procedure for two different PTAs: one mimicking the properties of the NANOGrav 15-year data set~\cite{NANOGrav:2023hde}, and another designed to resemble the expected properties of the upcoming third data release (DR3) of the IPTA collaboration. Compared to previous approaches, this way of generating null distributions automatically accounts for the correlations between the estimators of the cross-correlation coefficients, as well as for the non-Gaussianities in their distribution. Following this procedure, we estimate the null distribution for all the combinations of map parametrization and detection statistic considered in this work and summarized in Table~\ref{tab:summary}. In Fig.~\ref{fig:snr_null} and Fig.~\ref{fig:cl_null_dist}, we report some of the null distributions and upper limits derived in this way. Specifically, in Fig.~\ref{fig:snr_null} we show the null distribution for the anisotropic SNR of maps reconstructed using a square-root spherical harmonic (left panel) and pixel (central panel) basis, as well as the max-pixel SNR for maps reconstructed using the radiometer basis (right panel). In Fig.~\ref{fig:cl_null_dist}, we show the null distribution for the $C_\ell$ coefficients of maps reconstructed using the square-root spherical harmonic basis.

The final step of the detection pipeline consists of using the estimated null distributions to quantify the statistical significance of any deviation from isotropy observed in the reconstructed sky map.  For any map, we can compute the value of the detection statistic of interest. In general, the detection statistic is not a single scalar quantity, but rather a set of values - one associated with each pixel, frequency bin, or spherical harmonic coefficient, depending on the map parametrization and detection statistic used in the analysis. Let's start discussing the simplest case, one in which the detection statistic is a single scalar quantity. This is the case for both the anisotropic and the max-pixel radiometer SNR for broadband searches. In this case, the statistical significance of the measured SNR can be quantified by computing its $p$-value. Practically, the $p$-value is estimated by calculating the fraction of SNRs in the null distribution that have a value equal to or larger than the measured SNR. In this work, we classify any map that yields a detection statistic with a $p$-value$<3\times 10^{-3}$ as a detection of deviation from isotropy. 

For cases in which the detection statistic is not a single scalar quantity—such as for the Radiometer SNR map or for any detection statistic computed in per-frequency searches—the procedure is more involved. In these cases, we first compute a \emph{local $p$-value} for each component of the detection statistic (e.g., each pixel or frequency bin), following the same principle used in the scalar case: the local $p$-value is defined as the fraction of realizations in the null distribution for which that specific detection statistic component exceeds the observed value. This collection of local $p$-values is then translated to a \emph{global $p$-value} by computing the fraction of realizations in the null distribution for which at least one local $p$-value is smaller than the smallest local $p$-value observed in the measured detection statistic. We then classify as detection of deviation from isotropy any map that yields a detection statistic with a global $p$-value$<3\times 10^{-3}$. One example of the results  obtained with this procedure is shown in Fig.~\ref{fig:map_upper}, where we report the radiometer SNR values corresponding to a global $p$-value$=3\times 10^{-3}$. 

\begin{figure}[t!]
\centering
    \includegraphics[width=\textwidth]{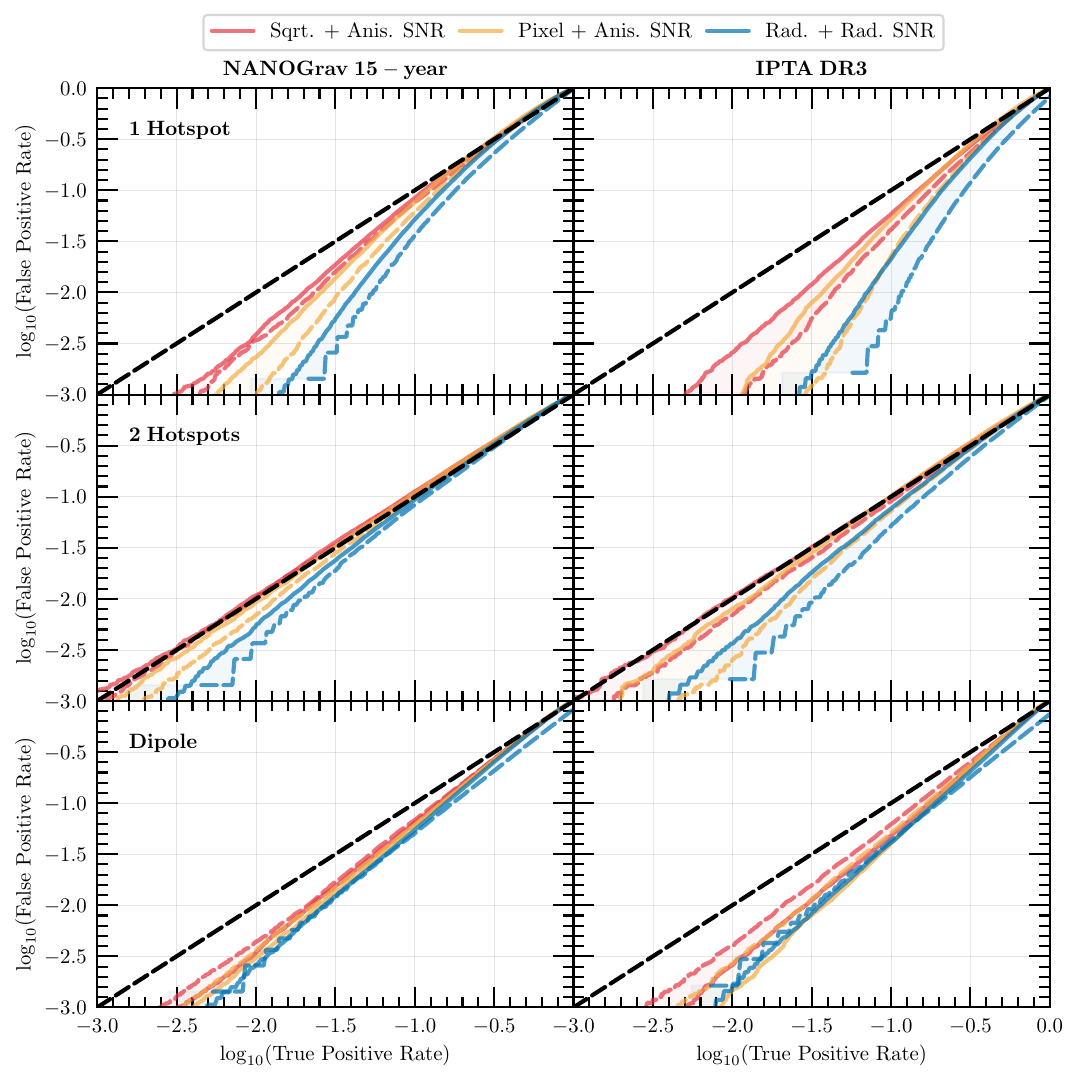}
    \caption{In this plot, for each map parametrization, we report the ROC curves for the best performing detection statistic (according to the results reported in Table~\ref{tab:summary}) and for different GWB anisotropies: single GWB hotspot (upper panels), two GWB hotspots (middle panels), and GWB dipole (lower panels). Specifically, on the x-axis, we report the global true positive rate, while on the y-axis we report the false positive rate of these different detection methods; both for the broadband searches (solid lines) and the per-frequency searches (dashed lines). The left (right) panels show the results for a PTA data set with similar noise properties to the NANOGrav 15-year (IPTA DR3) data set. The black dashed line is the ROC curve for the null distribution. The lines that are further from the black dashed line indicate the best-performing search strategies.}
    \label{fig:roc}
\end{figure}

\begin{table}[t]
    \centering
    \medskip
\begin{adjustbox}{max width=\textwidth}
\begin{NiceTabular}{@{}cc ccc ccc@{}}[cell-space-limits = 5pt, columns-width = 58pt]

\toprule
    \Block{2-1}{Map Basis} & \Block{2-1}{Detection Statistic} & \Block{1-3}{Detection Probability -- NG15} &&& \Block{1-3}{Detection Probability -- IPTA DR3} \\ 
\cmidrule(lr){3-5} \cmidrule(lr){6-8}
        & & 1 hotspot & 2 hotspots & Dipole & 1 hotspot & 2 hotspots & Dipole \\ 
\midrule
\Block{2-1}{Radiometer}        & Radiometer SNR     &  {\color{narrow}2.7\%} \quad {\color{broad}2.4\%} &{\color{narrow}0.73\%} \quad {\color{broad}0.61\%}&{\color{narrow}0.86\%} \quad {\color{broad}1.1\%}
                               &  {\color{narrow}7.2\%} \quad {\color{broad}4.1\%} &{\color{narrow}1.4\%} \quad {\color{broad}0.8\%}&{\color{narrow}1.1\%} \quad {\color{broad}1.5\%}\\
                               & Max Radiometer SNR       &  {\color{narrow}2.7\%} \quad {\color{broad}2.4\%} &{\color{narrow}0.70\%} \quad {\color{broad}0.48\%}&{\color{narrow}0.88\%} \quad {\color{broad}0.92\%}
                               &  {\color{narrow}6.7\%} \quad {\color{broad}3.1\%} &{\color{narrow}1.3\%} \quad {\color{broad}0.54\%}&{\color{narrow}0.96\%} \quad {\color{broad}1.1\%}\\
\midrule
\Block{1-1}{Pixel}     & Anisotropic SNR       & {\color{narrow}1.8\%} \quad {\color{broad}1.2\%}   &{\color{narrow}0.49\%} \quad {\color{broad}0.35\%}&{\color{narrow}0.86\%} \quad {\color{broad}1.0\%}
                       & {\color{narrow}4.8\%} \quad {\color{broad}1.9\%}   &{\color{narrow}1.1\%} \quad {\color{broad}0.43\%}&{\color{narrow}1.1\%} \quad {\color{broad}1.7\%}\\
\midrule
\Block{2-1}{Sqrt Sph. Har.}   & Anisotropic SNR     &  {\color{narrow}0.91\%} \quad {\color{broad}0.84\%} &{\color{narrow}0.30\%} \quad {\color{broad}0.28\%}&{\color{narrow}0.70\%} \quad {\color{broad}0.90\%}
                               &  {\color{narrow}2.3\%} \quad {\color{broad}1.1\%} &{\color{narrow}0.47\%} \quad {\color{broad}0.31\%}&{\color{narrow}0.76\%} \quad {\color{broad}1.1\%}\\
                               & $C_\ell$       &  {\color{narrow}0.34\%} \quad {\color{broad}0.34\%} &{\color{narrow}0.30\%} \quad {\color{broad}0.29\%}&{\color{narrow}0.31\%} \quad {\color{broad}0.41\%}
                               &  {\color{narrow}0.45\%} \quad {\color{broad}0.57\%} &{\color{narrow}0.32\%} \quad {\color{broad}0.36\%}&{\color{narrow}0.35\%} \quad {\color{broad}0.61\%}\\
\bottomrule  
\end{NiceTabular}
\end{adjustbox}
\caption{Comparison of detection probabilities for different combinations of map parametrization (first column) and detection statistic (second column). The values in blue (red) indicate detection probabilities for the per-frequency (broadband) search. Results are shown for two PTA data sets — one with noise characteristics similar to the NANOGrav 15-year data set, and one approximating the upcoming IPTA DR3. Detection probabilities are evaluated for three types of GWB anisotropies: a single hotspot, two hotspots, and a broadband dipole.}
\label{tab:summary} 
\end{table}

\section{Benchmarking anisotropies search strategies}\label{sec:benchmarks}
In this section, we benchmark several combinations of map reconstruction methods and detection statistics by estimating their current and future abilities to detect GWB anisotropies with different topologies. Specifically, we will consider GWB skies with one (Sec.~\ref{subsec:1hot}) or two (Sec.~\ref{subsec:2hot}) bright hotspots, as well as skies with large-scale anisotropies (Sec.~\ref{subsec:dipole}).  

\subsection{GWB Hotspot}\label{subsec:1hot}
The first scenario we use to benchmark the methods discussed in the previous sections consists of a GWB sky featuring a single, localized hotspot in an otherwise isotropic background. This scenario mimics the signal expected from a single, loud supermassive black hole binary dominating the GWB power in a narrow frequency window. Specifically, we use \texttt{pta\_replicator} to create $\sim 10^5$ realizations of the TOAs (and associated timing residuals) produced by a single source plus an isotropic GWB (in addition to white and red noise processes). The frequency of the single source is set to coincide with the center of the second frequency bin of a PTA with an observing time of $T_{\rm obs}=15\,{\rm yr}$ (i.e. $f=2/15\,{\rm yr}$) while its amplitude is chosen such that the single source contributes to 80\% of the GW power in the second frequency bin. The amplitude of the GWB in that frequency bin is reduced accordingly, such that the power spectrum of the GWB remains a power law without any bump at the hotspot frequency. The position, phase, and polarization of the single source are randomly assigned in each realization (see App.~\ref{app:mock_data} for more details on the procedure used to generate the mock data). For each of these mock data sets, we derive a set of estimators for the cross-correlation coefficients using \texttt{defiant}~\cite{Gersbach:2024hcc}, then pass these set of cross-correlations to reconstruct the GWB sky maps --using one of the map parametrizations discussed in Sec.~\ref{subsec:sky}-- and calculate the value of one of the detection statistics discussed in Sec.~\ref{subsec:snr}. The values of the detection statistics derived in this way are then compared with the null distributions discussed in Sec.~\ref{subsec:snr} to quantify the evidence for any deviation from the isotropic assumption. All the realizations that lead to a value of the detection statistic with a global $p$-value$<3\times10^{-3}$ are classified as detections of deviation from the isotropic assumption. We then estimate the probability of detecting the anisotropic signal produced by a single hotspot by computing the ratio between the number of detections and the total number of mock data sets. 

\begin{figure}[t!]
\centering
    \textbf{Radiometer basis, broadband}
    \includegraphics{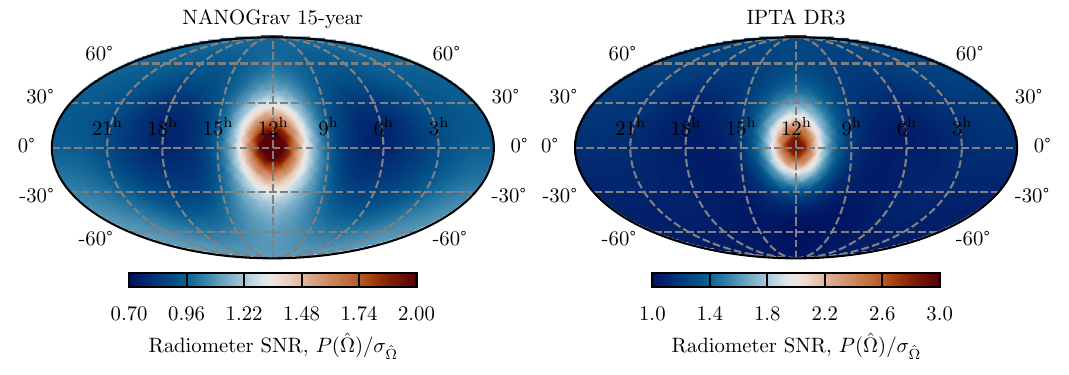}\vspace{1em}
    \textbf{Pixel basis, broadband}
    \includegraphics{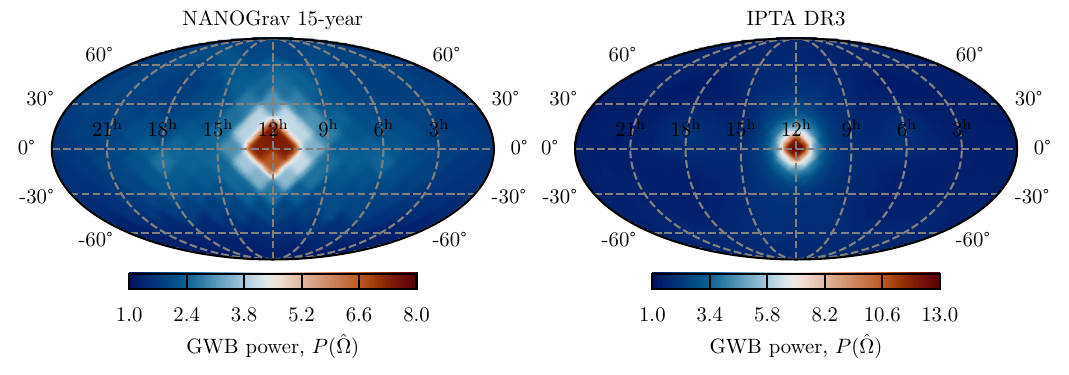}\vspace{1em}
    \textbf{Square-root spherical harmonics basis, broadband}
    \includegraphics{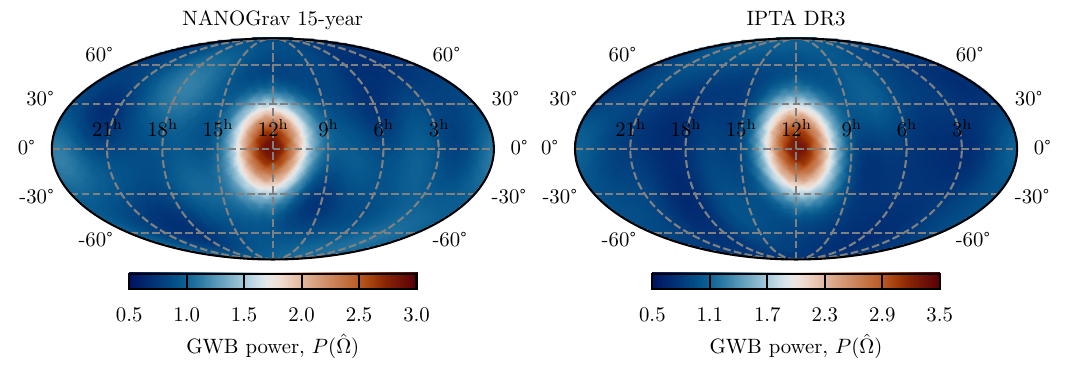}
    \caption{Average of reconstructed GWB maps (using a broadband search) for the single hotspots benchmark. Maps are rotated to place the hotspot at the map center before averaging. \emph{Top panels:} Radiometer SNR maps. \emph{Middle panels:} Pixel basis maps. \emph{Bottom panels:} Square-root spherical harmonics basis maps. Left (right) panels use PTA data with noise similar to the NANOGrav 15-year (IPTA DR3) data set.}
    \label{fig:hot_map}
\end{figure}

\begin{figure}[t!]
\centering
    \textbf{Radiometer basis, broadband}
    \includegraphics{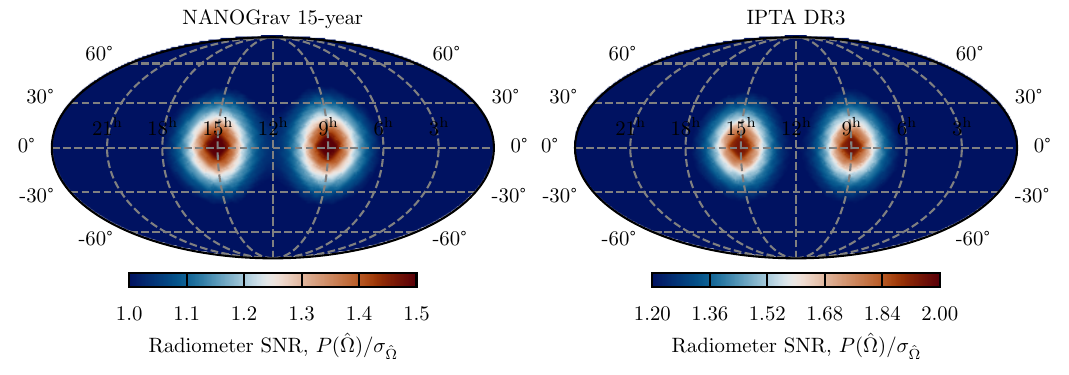}\vspace{1.em}
    \textbf{Pixel basis, broadband}
    \includegraphics{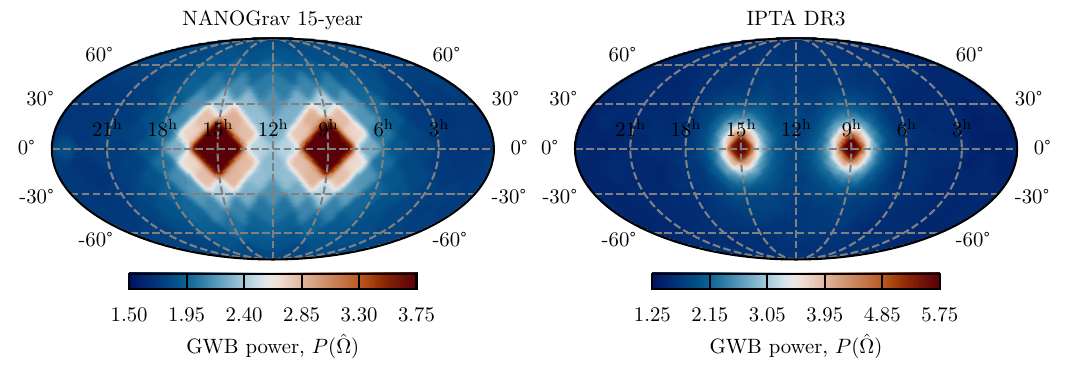}\vspace{1.em}
    \textbf{Square-root spherical harmonics basis, broadband}
    \includegraphics{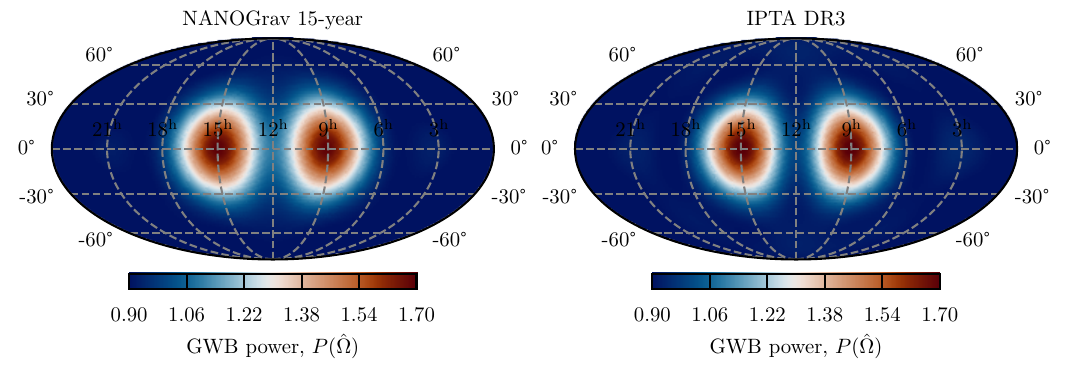}
    \caption{Average of reconstructed GWB maps (using a broadband search) for the two-hotspots benchmark. Before taking the average, maps are rotated to place the hotspots on the equatorial line and center them around the map center. \emph{Top panels:} Radiometer SNR maps. \emph{Middle panels:} Pixel basis maps. \emph{Bottom panels:} Square-root spherical harmonics basis maps. Left (right) panels use PTA data with noise similar to the NANOGrav 15-year (IPTA DR3) data set.}
    \label{fig:2hot_map}
\end{figure}

\begin{figure}[h!]
\centering
    \textbf{Radiometer basis, broadband}
    \includegraphics{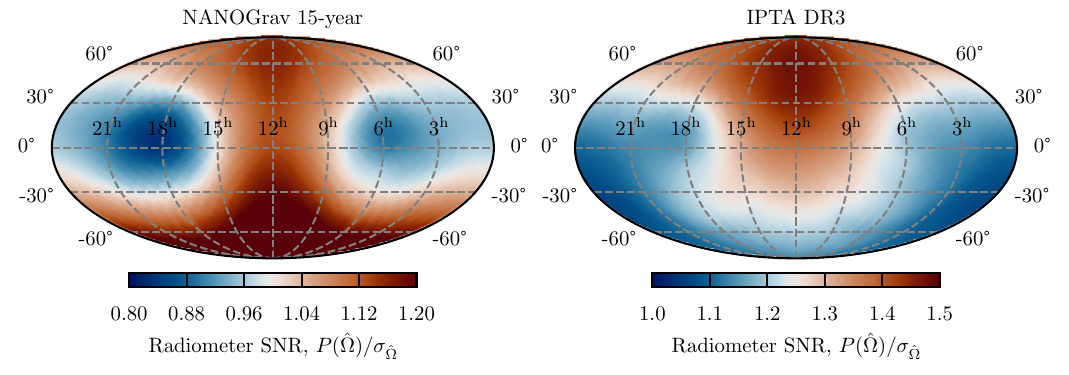}\vspace{1em}
    \textbf{Pixel basis, broadband}
    \includegraphics{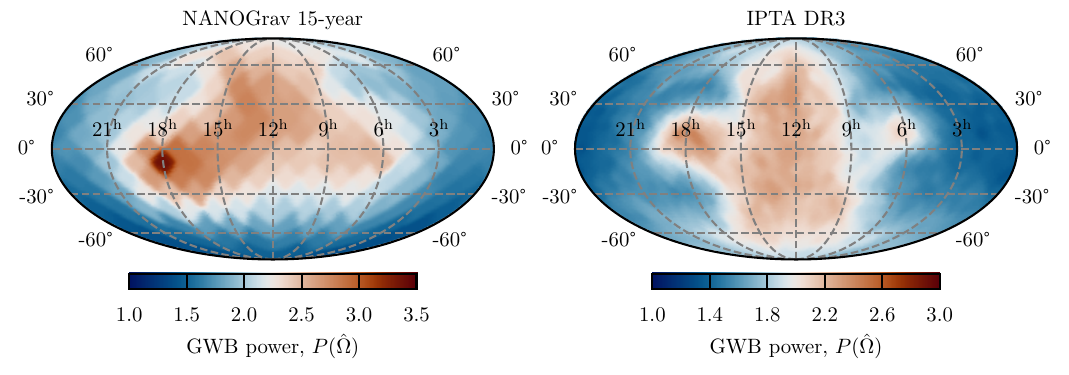}\vspace{1em}
    \textbf{Square-root spherical harmonics basis, broadband}
    \includegraphics{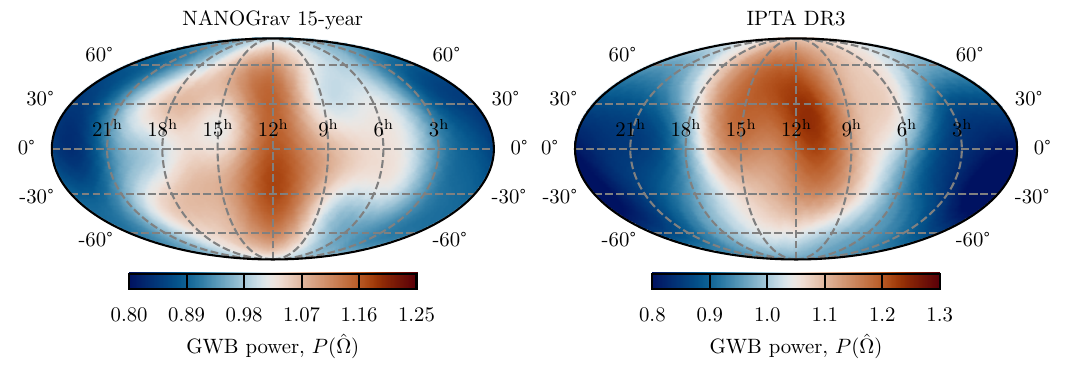}
    \caption{Average of reconstructed GWB for the dipole benchmark. Maps are rotated to place the bright side of the dipole at the map center before averaging. \emph{Top panels:} Radiometer SNR maps. \emph{Middle panels:} Pixel basis maps. \emph{Bottom panels:} Square-root spherical harmonics basis maps. Left (right) panels use PTA data with noise similar to the NANOGrav 15-year (IPTA DR3) data set.}
    \label{fig:dip_map}
\end{figure}

The results of this analysis are summarized in Table~\ref{tab:summary} and Fig.~\ref{fig:roc}. In Table~\ref{tab:summary}, we report the detection probabilities for the different detection methods. We see that per-frequency searches perform better than broadband searches for all combinations of map parametrization and detection statistics, which is expected since the signal we are looking for is localized in frequency space. Across the different map parameterizations, the radiometer basis is the best performing one for maps with a single GWB hotspot, independently from the detection statistic used (with both the radiometer SNR and max pixel SNR leading to similar detection rates). Again, this aligns with our expectations as we are considering a single, well-localized source that dominates the GWB in the frequency bin. However, even for the best performing search strategy, detection rates do not reach $10\%$ even for a PTA data set with noise properties similar to the ones of the upcoming IPTA DR3, indicating that even extremely bright hotspots will be challenging to detect with near-future PTA sensitivities. 

To assess the accuracy of the map reconstruction, in Fig.~\ref{fig:hot_map} (see Fig.~\ref{fig:hot_map_pf} at the end of this manuscript for the results of the per-frequency search), we show the average of the maps reconstructed using the different map parametrizations discussed in this work. Before performing the average, each map is rotated such that the location of the injected hotspot falls at the center of the map (i.e. it has a $0^\circ$ declination angle, and a $180^\circ$ right ascension). From this figure, we can see how all map parametrizations allow, on average, to accurately localize the location of the hotspot, with the pixel basis providing the most accurate sky localization.

\subsection{Two GWB Hotspots}\label{subsec:2hot}

The second kind of GWB anisotropy that we consider is the one produced by two localized hotspots, each contributing to $40\%$ of the total GWB power in the second frequency bin (i.e. $f=2/T_{\rm obs}$). As for the single hotspots case, the sky position, phase, and polarization of the two sources are randomly assigned in each realization, with the constraint of setting the angular distance between the sources to $\pi/2$ (see App.~\ref{app:mock_data} for more details on the procedure used to generate the mock data). The choice of placing both the hotspots in the same frequency bin is motivated by the idea of maximizing interference between their signals and studying how this impacts both detection forecasts and map reconstruction. 

To estimate detection probabilities, we generate $\sim10^5$ realizations of pulsar timing data in which we inject the GW signals of two hotspots in addition to an otherwise isotropic GWB, and then follow the same procedure already outlined for the single hotspot case. The results of this analysis are summarized in Table~\ref{tab:summary} and Fig.~\ref{fig:roc}. Despite the signal being no longer dominated by a single bright source, we find that the radiometer basis is still the best performing map parametrization, independent of the detection statistic that it is used with. We also see that frequency-resolved searches continue to outperform broadband ones. However, detection probabilities drop by roughly a factor of five compared to the single hotspots scenario, reinforcing the idea that realistic GWB anisotropies produced by a collection of bright SMBHB binaries might be challenging to detect in the near future. Averages of GWB maps reconstructed with different map parametrizations are shown in Fig.~\ref{fig:2hot_map} (see Fig.~\ref{fig:2hot_map_pf} at the end of this manuscript for the results of the per-frequency search).  Before performing the average, each map is rotated such that the locations of the two injected hotspots fall on the equatorial axis, and their mid-point is aligned with the center of the maps. As we can see from this figure, the hotspots locations are --on average-- accurately reconstructed with all three map parametrizations. For the NANOGrav-like case, the best localization is achieved with the radiometer reconstruction. For the IPTA-like data sets, the hotspots are most accurately localized with the pixel maps, as this map parametrization allows us to use a higher resolution without facing numerical problems.

\subsection{Detecting Large-Scale Anisotropies}\label{subsec:dipole}
The last benchmark that we consider is a GWB dipole. In contrast to the two previous benchmarks, where anisotropies were localized both in position and frequency space, in this scenario, the GWB anisotropies will be on large scales and broadband. Specifically, we consider a GWB with spherical harmonics coefficients of $C_1 = C_0/9$ and $C_{\ell >1}=0$. This is the largest allowed value of $C_1$ that still allows for preservation of positivity of the GWB power over the entire sky while keeping higher multipoles set to zero. 

To estimate detection probabilities, we generate $\sim10^5$ realizations of a PTA data set in which we inject the signal of this anisotropic GWB. For each realization, the direction of the dipole is randomly drawn. We then follow the same procedure used for the two previous benchmarks to derive detection probabilities. The results of this analysis are reported in Table~\ref{tab:summary} and Fig.~\ref{fig:roc}. We find that all combinations of map parametrization and detection statistics have similar performance, with broadband searches performing slightly better than per-frequency ones. However, even for an IPTA DR3-like data set, detection probabilities never exceed the $\sim1\%$ level.

Averages of GWB maps reconstructed with different map parametrizations are shown in Fig.~\ref{fig:dip_map} (see Fig.~\ref{fig:dip_map_pf} at the end of this manuscript for the results of the per-frequency search).  Before performing the average, all the maps are rotated such that the location of the dipole with excess power is located at the center of the sky map. From this figure, we can see how accurately reconstructing the injected dipole is more challenging compared to the case of localized hotspots. The quality of the map reconstruction is particularly poor for the radiometer maps, which is expected given that the radiometer parametrization assumes all the power is confined to a single pixel—an assumption that breaks down in the presence of large-scale anisotropies.

\section{Fundamental limits of PTA anisotropy searches}\label{sec:cosmic_variance}

The goal of this section is to address the following question: Consider an ideal PTA in which the TOAs can be measured with arbitrary precision, and the only contribution to the timing residuals comes from the GWB (i.e., there is no white noise or intrinsic pulsar red noise). What is the level of GWB anisotropies that can be detected by such an ideal PTA? This question can also be phrased as: What is the fundamental limit to anisotropy detection posed by cosmic variance? Indeed, even in the absence of noise, the estimators of the pulsars' cross-correlations coefficients will have a non-vanishing variance --also known as cosmic variance-- around their true value~\cite{Allen:2022dzg}. This can be seen explicitly from the fact that Eq.~\eqref{eq:rho_cov} remains finite even when all noise sources are set to zero due to GWB self-noise. This intrinsic variance will, in turn, set a fundamental limit to the magnitude of deviation from isotropy that can be detected by PTAs. The goal of this section is to find this fundamental limit and study how this scales with observing time and the number of pulsars in the PTA. 

Analytical estimates for the smallest detectable GWB dipole were previously derived in Ref.~\cite{Hotinli:2019tpc} using bipolar spherical harmonics. There, it was shown that---in the signal-dominated regime---the minimum detectable dipole anisotropy in a PTA scales as $C_1 \propto 1/N_p$, with an additional $1/T_{\rm obs}$ scaling in the case of broadband anisotropies. In this work, we test whether the anisotropy search strategies applied in realistic PTA analyses are able to reach this fundamental limit. Moreover, we extend the analysis to GWB anisotropies different from a simple dipole.
\begin{figure}[t!]
\centering
    \includegraphics{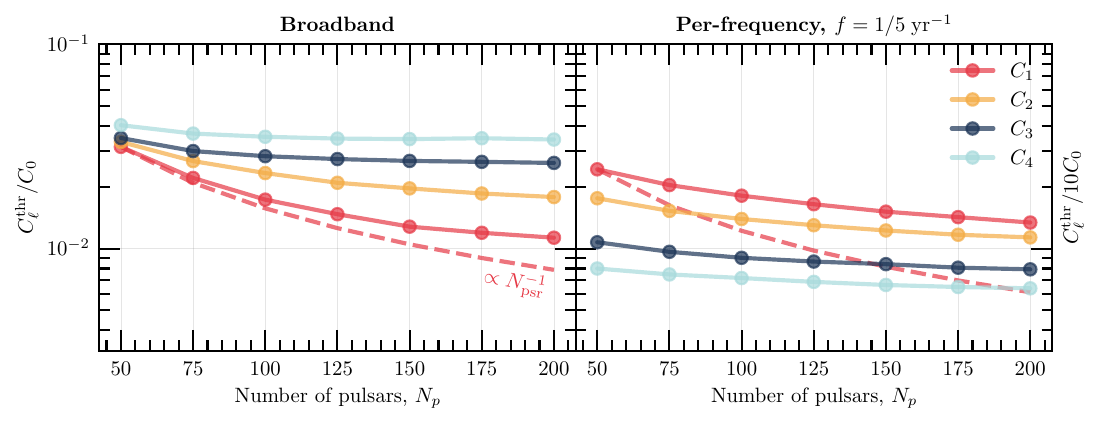}\\ 
    \includegraphics{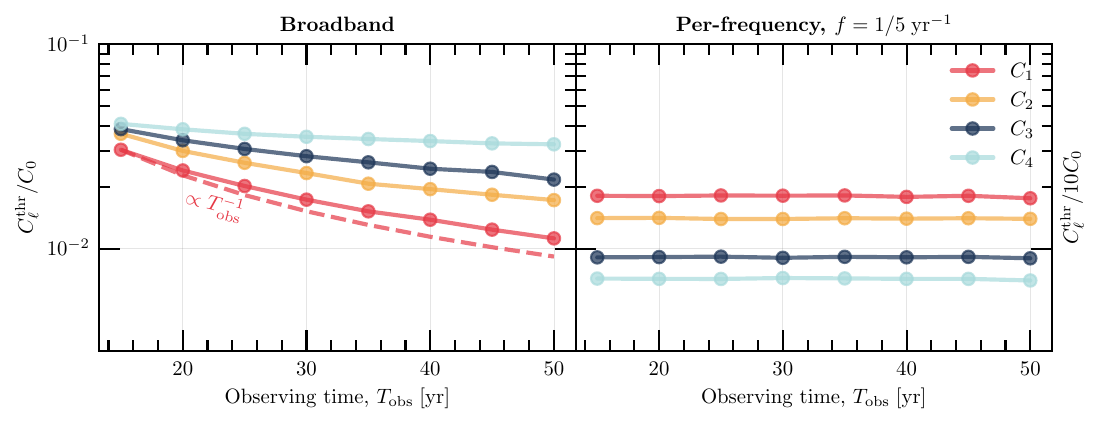}
    \caption{Upper limits on the $C_\ell$ in the square-root spherical harmonics basis for ideal PTA configurations. In the left panel, we show the results of a broadband search, while in the right panels we show the upper limits placed by a per-frequency search on the anisotropies in the frequency bin $f=0.2\,{\rm yr}^{-1}$. In the upper panels, we show the evolution of these limits depending on the number of pulsars in the PTA. In the lower panels, we show the evolution with the observation baseline. For the varied pulsar number case, the observation baseline is fixed to $T_{\rm obs} = 30\rm{yr}$.  For the varied observing baseline case, we fix the number of pulsars to $N_p=100$. The dashed red lines show the scaling behaviours predicted in Refs.~\cite{Hotinli:2019tpc} for the dipole upper limits.}
    \label{fig:ideal_sqrtharm}
\end{figure}

\begin{figure}[t!]
\centering
    \includegraphics{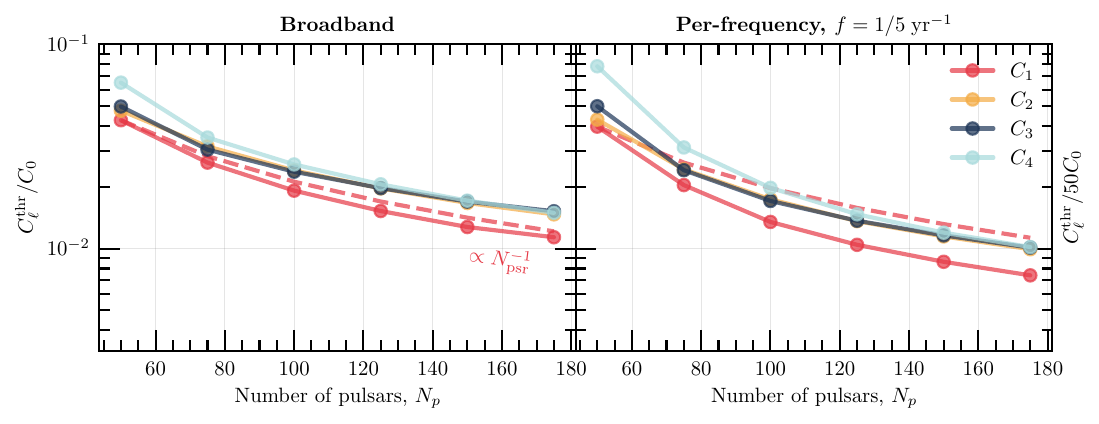}\\ 
    \includegraphics{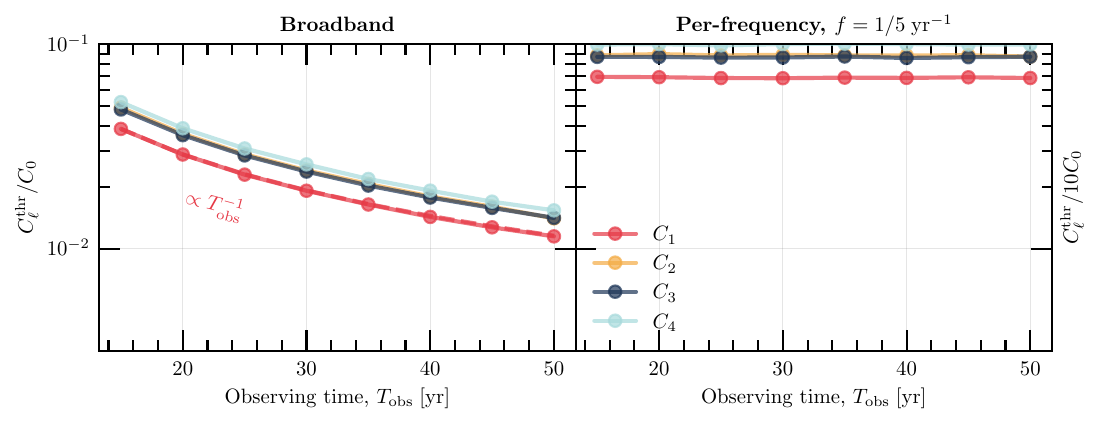}
    \caption{Same as Fig.~\ref{fig:ideal_sqrtharm}, but for the linear spherical harmonics basis.}
    \label{fig:ideal_linear}
\end{figure}

\begin{figure}[t!]
\centering
    \includegraphics{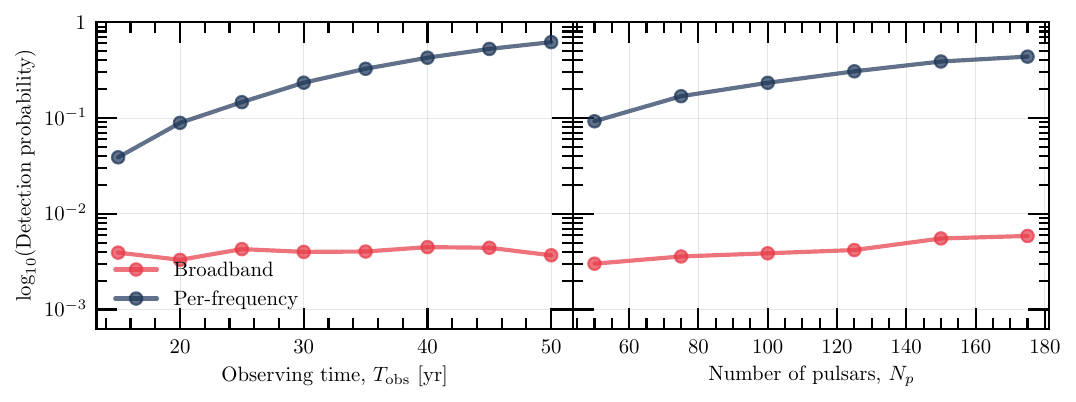}\\
    \caption{Detection probabilities for a point source with an ideal PTA configuration using the maximum radiometer SNR. The red (blue) lines show the results obtained with the broadband (per-frequency) reconstruction. The left (right) panel shows the probabilities as a function of the observation baseline (number of pulsars in the PTA). For the varied pulsar number case, the observation baseline is fixed to $T_{\rm obs} = 30\rm{yr}$, the PTA configuration for the varied baseline case consists of 100 pulsars.}
    \label{fig:ideal_rad}
\end{figure}

\subsection{Large-scale anisotropies}
We begin by deriving the upper limits that an ideal, noise-free PTA could place on large-scale, broadband anisotropies.
To do this, we repeat the square-root spherical harmonics map reconstruction described in Sec.~\ref{sec:gwb_signal}, but this time for an idealized PTA, whose data stream contains only the GWB signal and no other noise source. Specifically, we consider different PTA configurations with different combinations of pulsars (that we assume to be homogeneously distributed in the sky) and observing time. For each of these configurations, we generate 28800 mock PTA data sets in which we inject the signal of an isotropic GWB. We then perform a full search for anisotropies in these mock data sets using a square-root map parametrization, from which we derive null distributions and set upper limits on the $C_\ell$ defined in Eq.~\eqref{eq:Cl}.

We report the results of this analysis in  Fig.~\ref{fig:ideal_sqrtharm}, where we show upper limits on the $C_\ell$ in the square-root spherical harmonics basis as functions of the number of pulsars and observation time.\footnote{These upper limits correspond to $C_\ell$ with a local $p$-value of $p=3\times10^{-3}$.}
We find that broadband searches get more and more sensitive as we increase the number of pulsars and the observing time, indicating that the limitations introduced by cosmic variance could be arbitrarily reduced by increasing the number of pulsars or --for broadband signals-- the observing time. However, these results also suggest that --even in the absence of noise-- the limitation introduced by cosmic variance would make the detection of anisotropies with $C_\ell\lesssim10^{-2}$ highly challenging for any realistic PTA configuration. For example, the motion of the solar system is expected to imprint a kinematic dipole in GWB of cosmological origin. The amplitude of this dipole is of order $C_1/C_0\sim10^{-6}$. Therefore, from the above discussion, we can conclude that detecting such kinematic anisotropies would be practically impossible for any realistic PTA configuration. The situation is even less promising for per-frequency searches. In this case, the upper limits are dominated by the priors indirectly placed on the $C_\ell$ by the square-root spherical harmonic basis. These prior effects are the reason why the constraints seem to be stronger for higher multipoles, as can be seen in the right panels of Fig.~\ref{fig:ideal_sqrtharm}, where we report the upper limits on the spherical harmonic coefficients of the anisotropies in the frequency bin $f=0.2\,{\rm yr}^{-1}$. 

In our results, we also find deviations from the expected $1/N_p T_{\rm obs}$ scaling for the $C_\ell$ upper limits of broadband searches, with the magnitude of these deviations becoming more and more pronounced for higher multipoles. We find a similar behavior for the per-frequency search, where the upper limits for all the multiples --while improving when increasing the number of pulsars-- show deviations from the expected $1/N_p$ scaling. We suspect these deviations to arise from the priors implicitly placed on the $C_\ell$ in the square-root basis. Therefore, to isolate the effect of these priors, we repeat this analysis using the linear spherical harmonics basis. 
For this basis, we directly parametrize the GWB power using the spherical harmonic functions $Y_{\ell m}$. The power map is then given by a linear combination of these functions weighted by their coefficients $c_{\ell m}$:
\begin{equation}
    P_p=\sum_{\ell=0}^{\ell_{\rm max}}\sum_{m=-\ell}^{m=\ell}c_{\ell m}Y_{\ell m}(\hat\Omega_p)\,.
\end{equation}
As for the case of the square-root spherical harmonics analysis, we set $\ell_{\rm max}=7$.
The maximum-likelihood values for the $c_{\ell m}$, $\hat{\bm{c}}$, can be calculated analytically using the same approach as for the radiometer basis:
\begin{equation}
    \hat{\bm{c}} = \bm{M}^{-1}\bm{X},
\end{equation} 
where we use a spherical harmonics version of the antenna response function, 
\begin{equation}
    R_{(\ell m)(ab)} = \frac{3}{2 N_{\rm pix}}\sum_p \ Y_{lm}(\hat\Omega_p)\left[F_{ap}^+F_{bp}^++F_{ap}^\times F_{bp}^\times\right],
\end{equation}
to define the Fisher matrix and the dirty map.

The cosmic-variance-limited $C_\ell$ upper limits are shown in Fig.~\ref{fig:ideal_linear}. In this case, the upper limits almost exactly follow the scaling relations derived in Ref.~\cite{Hotinli:2019tpc}, confirming that the deviations observed for the square-root basis analysis were mostly due to priors effects.

\subsection{GWB hotspot}
In this section, we investigate the limitations cosmic variance poses for the detection of GWB hotspots. We do this by generating 28800 realizations of a PTA data set in which we inject the signal of a GWB containing a bright hotspot at a frequency of $f=1/5~\rm{yr}^{-1}$ and with a GW strain of $h_s\simeq 1.4\times10^{-15}$, corresponding to $50\%$ of the GW power in the 10-th bin of a PTA with $T_{\rm obs}=50~\rm{yr}$.\footnote{Notice that, since we are keeping the frequency hotspot signal fixed, PTAs with different observing times will observe the hotspot signal in difference frequency bins. Moreover, since we are keeping the physical amplitude of the signal fixed, the contribution of the hotspot to the total GWB power in that bin will scale as $T_{\rm obs}$, given that the bin size will get smaller and smaller for longer observing times, while the amplitude of the hotspot remains constant.} We then perform a search for GWB anisotropies in these mock data sets using the radiometer map parametrization and maximum pixel SNR detection statistic that we calibrate by repeating this procedure on a set of 28800 mock data sets, which contain the signal of an isotropic GWB. 

The results of this analysis are shown in Fig.~\ref{fig:ideal_rad}, where we report anisotropy detection probabilities in these idealized PTAs as a function of both time and number of pulsars. For per-frequency searches, detection probabilities increase both as a function of observing time and number of pulsars in the array, indicating that the limitations introduced by cosmic variance can be arbitrarily reduced by increasing the size of the PTA or the total observing time. 
For broadband searches, detection probabilities remain constant around $3\times10^{-3}$ (which is the false positive rate chosen for our detection threshold) when increasing the observing time. This is to be expected as lower frequencies with higher GWB amplitudes start to be accessible for longer observing times, leading to a relative suppression of high-frequency signals in a broadband search. When increasing the number of pulsars in the array, detection probabilities slightly increase also for broadband searches, but at a slower rate compared to the per-frequency search, further emphasizing the importance of frequency-resolved anisotropy searches.

\section{Conclusions}\label{sec:conclusions}
Several techniques have been proposed to search for potential deviations from isotropy in the GWB observed by multiple PTA collaborations. In this work, we perform a systematic comparison of the most advanced of these techniques by applying them to realistic mock PTA data sets. Our objective is to understand which search strategies are the best suited to detect anisotropies of different topologies and identify their fundamental limitations. The three main results of this work are the following:

\begin{enumerate}
    \item The radiometer map parametrization, combined with the radiometer SNR detection statistic, seems to be the best performing search strategy across the different anisotropic signals considered in this work. The pixel basis, combined with the anisotropic SNR, is a close second, while the square-root spherical harmonic parametrization seems to constantly underperform when compared with these other search strategies. \\
    Moreover, for all the detection methods that we tested, per-frequency searches greatly outperform broadband searches when it comes to narrowband anisotropic features, like the ones expected from a population of SMBHB.

    \item Even the best performing search strategy tested in this work, combined with a PTA with the noise properties of the upcoming IPTA DR3, struggles to detect the large anisotropic signals considered in the work. For example, detection probabilities for a GWB hotspot contributing to $80\%$ of the power in the second frequency bin never exceed 10$\%$ in our study. This consolidates some of the findings of previous works~\cite{Lemke:2024cdu}, and emphasizes how challenging it will be to detect SMBHB-generated anisotropies. 

    \item For a given PTA setup, cosmic variance sets a lower limit to the level of GWB anisotropies that can be detected with PTA observations. 
    These lower limits can be arbitrarily reduced by increasing the observing time and the number of pulsars in the array. However, for a realistic PTA, it will be challenging to push this fundamental limit below $C_{\ell=1}/C_{\ell=0}\lesssim10^{-2}$. This fundamental limit will make the detection of intrinsic or kinematic anisotropies in a primordial GWB practically impossible. \\
\end{enumerate}

While in this work we tested some of the most advanced anisotropy search strategies currently proposed for PTA analyses, these strategies are still based on several assumptions that could be lifted to improve the sensitivity of the search. Here we list some of these assumptions:
\begin{itemize}
    \item The likelihood that is used to reconstruct the GWB sky maps, given in Eq.~\eqref{eq:likelihood}, assumes that the estimators for the cross-correlation coefficients, $\hat\rho$, are Gaussian distributed. In general, this is not true, and deviations from Gaussianity will become more prominent with lower noise levels. 
    
    \item We are assuming the GWB is unpolarized, but if we have a hotspot produced by a loud binary, we expect the signal to be strongly polarized. Map-making techniques that account for this effect could improve sensitivity, see for example Ref.~\cite{Jow:2025uut}.
    
    \item The construction of the pair covariance matrix, $\bm{\Sigma}$, entering in the likelihood used to estimate cross-correlation coefficients requires knowledge of both the GWB amplitude, and the pulsars' cross-correlation coefficients. This implies that a choice of the cross-correlation coefficients is needed to construct an estimator for this same quantity. In current searches, HD cross-correlations are used to build an estimator for both the GWB amplitude and the cross-correlations. However, for anisotropic skies, this is clearly an unjustified assumption.  A more robust strategy could use an iterative approach, in which one begins by constructing an estimator based on the HD correlations, uses the resulting cross-correlation estimates to update $\bm{\Sigma}$, and then repeats the process until convergence.
    
    \item The response function used in anisotropy searches, see Eq.~\eqref{eq:antenna_response}, neglects the contribution of the pulsar term. While this is a justified assumption for isotropic GWB skies, for skies that present large and localized anisotropies, the contribution of the pulsar term can play an important role \cite{Boyle&Pen2012}. Including the pulsar term in the overlap reduction function would require knowing pulsar distances with a precision smaller than the wavelength of the GW of interest. In the absence of this information, the pulsar term could be included in a Bayesian analysis, and the uncertainty on the pulsars' distances numerically marginalized over.
\end{itemize}

While challenging, detection of GWB anisotropies remains one of the best ways to identify and characterize the source of the GWB observed in the nHz band. In this study, we have identified the most promising way to achieve this detection and highlighted some of the fundamental limitations of current anisotropy searches. However, a number of important questions still remain open, some of which the simulation framework developed in this work will help to address.
For example, extending the present analysis to Bayesian anisotropy searches and directly comparing their performance with that of frequentist methods would be highly valuable. Such a comparison could clarify whether Bayesian searches --by fully exploiting the information encoded in the timing residuals-- offer improved sensitivity to GWB anisotropies.
Moreover, it remains unclear whether the gravitational wave signal from loud binaries will first be detected through anisotropy searches or via searches for continuous GW signals. 
Finally, detailed anisotropy detection forecasts for SMBHB populations --which update the results of Ref.~\cite{Lemke:2024cdu} by including the impact of cosmic variance and by employing the new search strategies developed in Ref.~\cite{Gersbach:2025mhj} and benchmarked in this work-- are still needed to inform future astrophysical inference studies. 

\vspace{10pt}
\noindent While this work was being completed, Ref.~\cite{Domcke:2025esw} appeared on the arXiv. This work also investigates the fundamental limitations of PTA anisotropy searches introduced by cosmic variance. The authors find that these limitations can be arbitrarily reduced by increasing the number of pulsars in the PTA, consistent with what we have shown in this paper.

\section*{Acknowledgments}
This work was supported by the Deutsche Forschungsgemeinschaft under Germany’s Excellence Strategy - EXC 2121 Quantum Universe - 390833306. We thank Kyle A. Gersbach and Emiko C. Gardiner for useful discussions. This work used the Maxwell computational resources operated at Deutsches Elektronen-Synchrotron DESY, Hamburg (Germany).

\appendix
\section{Mock Data}\label{app:mock_data}
In this appendix, we provide additional information about the procedure used in this work to produce mock PTA data sets. 
\subsection{Signal Components}
All mock data sets used in this work have been generated using the \texttt{pta-replicator} package~\cite{pta_replicator} (or a modified version to allow for faster data simulation), a tool that builds upon \texttt{enterprise}~\cite{enterprise} and \texttt{pint}~\cite{Luo_2021, Susobhanan:2024gzf} to create mock TOAs and associated timing residuals. In this section, we describe all the assumptions and choices made in this procedure.

\noindent\textbf{Epoch-averaged TOAs:}
To make the data generation procedure more efficient and allow the generation of a large number of mock data sets,  we epoch-average the TOAs, accounting for measurement uncertainties, correlated white noise within observing epochs (ECORR), as well as EFAC and EQUAD. This approach reduces the number of TOAs in our simulated data sets significantly while still accounting for all relevant characteristics of the data set, such as irregular observation times, varying baselines, and noise properties. 
Specifically, we compute the combined TOA uncertainties as a noise-weighted average according to \cite{justin_ellis_2017_251456}:
\begin{equation}
    \sigma_{\rm TOA} = \frac{1}{\sqrt{\bm{1}^{T}\cdot \bm{N} \cdot\bm{1}}},
\end{equation}
where $\bm{N}$ is the white noise covariance matrix for all TOAs in the epoch as given by \eqref{eq:wn_matrix}. All white noise contributions are therefore
contained within the measurement uncertainties $\sigma_{\rm TOA}$ of the epoch-averaged TOAs.

\textbf{White Noise:}
The white noise contribution to the TOAs is simply drawn from a normal distribution centered around zero, with the standard deviation given by the TOA uncertainties:
\begin{equation}
    \delta t_{\rm WN} \sim \mathcal{N}(0,\sigma_{\rm TOA}^2).
\end{equation}
Since we are working with epoch-averaged TOAs, we do not need to account for correlated white noise (ECORR) or additional contributions (i.e. EFAC or EQUAD).

\textbf{Red Noise:}
The power spectra of pulsar-intrinsic noise processes are assumed to be a power law:
\begin{equation}
    \varphi_a(f)=\frac{A_a^2}{12\pi^2}\left(\frac{f}{{\rm yr}^{-1}}\right)^{-\gamma_a}\frac{{\rm yr}^3}{T_{\rm obs}}\,,
\end{equation}
where the amplitude, $A_a$, and slope, $\gamma_a$, of this power law are pulsar-dependent parameters.
Given this power spectrum, sine and cosine components, $\tilde a$, for the intrinsic red noise are drawn according $\tilde a(f)\sim\mathcal{N}(0,\varphi_{a}(f))$ for $f=1/T_{\rm psr}, 2/T_{\rm psr},...,30/T_{\rm psr}$, where $T_{\rm psr}$ is the individual observation baseline for each pulsar. 
The time delays then computed by multiplying these frequency domain components with the Fourier design matrix $\bm{F}$ as given in Eq.~\eqref{eq:fourier_design}, so that we obtain $\delta t_{\rm{RN}}=\bm{F}\tilde{a}$.

We do not account for other noise sources, such as chromatic events, DM noise, chromatic wind or deterministic effects, instead, we leave the study of their effects on anisotropy searches to future work.

\textbf{Gravitational Wave Background:}
The GWB includes contributions between $f_{\min}=1/(10T)$ and $f_{\rm max}=300/T$ in increments of $\Delta f=1/(10T)$. For each pulsar, the imaginary and random parts of the coefficients in frequency domain, $P_{a}(f)$, are drawn from a zero-mean normal distribution with variance given by a powerlaw spectrum,
\begin{equation}
    \Phi(f)=\frac{A_{\rm gw}^2}{12\pi^2}\left(\frac{f}{{\rm yr}^{-1}}\right)^{-\gamma_{\rm gw}}\frac{{\rm yr}^3}{T_{\rm obs}}\,,
\end{equation}
so that we have $P_{a}(f)\sim \mathcal{N}(0,\Phi(f))$.
To account for the spatial correlations between the residuals in the pulsars, the vector of frequency domain residuals is multiplied by the Cholesky decomposition $\bm{C}$ of the overlap reduction function matrix, $\bm{\Gamma}_{ab}=\bm{C}^T\bm{C}$: 
\begin{equation}
    P_{a,\rm corr}(f) = \sum_{b}\bm{C}_{ab}P_{b}(f).
\end{equation}
For an isotropic GWB, $\Gamma_{ab}$ is given by the Hellings-Downs curve.
The correlated Fourier coefficients are then translated into the time domain with a fast Fourier transform as implemented in \texttt{numpy} and interpolated onto the observation times.

We choose $\gamma_{\rm GW}=-2/3$, the expected characteristic strain spectrum for a GWB sourced by a population of supermassive black hole binaries that evolve by GW emission only. We set the amplitude to be $\rm{log}_{10}A_{\rm{yr}}=-14.67$, coinciding with the maximum likelihood value for a GWB with this characteristic strain spectrum in the NANOGrav 15-year data set~\cite{NANOGrav:2023gor}.

\textbf{Hotspot:}
For the hotspot data sets, we modify the power spectrum of the isotropic GWB component. As we are interested in investigating purely the effects of spatial anisotropies and not the effects of deviations from a pure power law spectrum, we reduce the power in the isotropic component by exactly the amount of power we later inject as a continuous signal.

We assume the GW frequency and amplitude to be constant across the observing time to obtain the most optimistic estimates.
The initial orbital phase $\phi_0$, the polarization $\psi$ and the sky position, characterized by $\theta_{\rm GW}$ and $\phi_{\rm{GW}}$, are drawn randomly from uniform distributions. 
Using $\hat{{\Omega}}=(-\sin\theta_{\rm GW}\cos\phi_{\rm GW},-\sin\theta_{\rm GW}\sin\phi_{\rm GW},-\cos\theta_{\rm GW})$ and the antenna response functions $F_{a}^{A}(\hat{\Omega})$ as defined in Eq.~\eqref{eq:ar}, we compute the timing residuals as 
\begin{equation}
\begin{split}
    \delta t_a(t) = \frac{h_c}{2\pi f_{\rm GW}^{3/2}T_{\rm obs}^{1/2}}\big[F_a^{+}(\hat{{\Omega}})[&\cos(2\psi)\sin(\phi_{0}+2\omega t_{p}(t))+\sin(2\psi)\cos(\phi_{0}+2\omega t_{p}(t))\\
    &-\cos(2\psi)\sin(\phi_{0}+2\omega t)-\sin(2\psi)\cos(\phi_{0}+2\omega t)] \\
    +(F_a^{\times}(\hat{{\Omega}})[&-\sin(2\psi)\sin(\phi_{0}+2\omega t_{p}(t))+\cos(2\psi)\cos(\phi_{0}+2\omega t_{p}(t))\\
    &+\sin(2\psi)\sin(\phi_{0}+2\omega t)-\cos(2\psi)\cos(\phi_{0}+2\omega t)])\big],
\end{split}
\end{equation}
where the characteristic strain, $h_c$, is related to the GW strain, $h_s$, by $h_c = h_s \sqrt{f/\Delta f}$, and we have defined the pulsar time $t_{p}(t) = (t-d_a(1-\cos \mu))$ with $\cos\mu=-\hat{{\Omega}}\cdot\hat{{p}}$ and $\hat{p}$ the unit vector pointing towards the pulsar. The pulsar distances, $d_a$, are drawn from a uniform distribution with $d_{\min}=0.5~\rm{kpc}$ and $d_{\min}=1.5~\rm{kpc}$. We have checked that the specific choice of pulsar distances does not influence our results, therefore we keep them fixed across realizations.

\paragraph*{\bf Large-scale anisotropies:}
To inject a dipole signal into the timing residuals we follow the same procedure as for the isotropic GWB, but using a modified ORF.
The maximally dipolar injection we consider in this work consists in taking non-zero values only for the $\ell = 0,1$ spherical harmonic coefficients of the background power map and imposing a positivity constraint on them. This translates into having, once we normalize $c_{00}=\sqrt{4\pi}$, 
\begin{equation}
    c_{1 -1} = - \sqrt{\frac{4 \pi}{3}} \sin{\theta_D} \sin{\phi_D}, \quad c_{1 0} = \sqrt{\frac{4 \pi}{3}} \cos{\theta_D}, \quad c_{1 1} = - \sqrt{\frac{4 \pi}{3}} \sin{\theta_D} \cos{\phi_D}. 
\end{equation}
Here $\theta_D$, $\phi_D$ identify the direction of the dipole in the sky and are drawn randomly from a uniform distribution for each sample of the mock data. 

\noindent From these coefficients the anisotropic overlap reduction function can be computed using the real spherical harmonics $Y_{\ell m}$ according to \cite{Mingarelli:2013dsa}
\begin{equation}
    \rho_{ab} = \sum_{\ell=0}^{\ell_{\rm max}}\sum_{m=-\ell}^{m=\ell}c_{\ell m} \, R_{(\ell m)(ab)}\,,
\end{equation}
where $R_{(\ell m)(ab)}$ is the ORF component corresponding to $Y_{\ell m}$ defined as:
\begin{equation}
    R_{(\ell m)(ab)} = \frac{3}{2}\int \frac{d \hat{{\Omega}}}{4\pi}\,Y_{lm}(\hat\Omega)\left[F_{a}^+(\hat{{\Omega}})F_{b}^+(\hat{{\Omega}})+F_{a}^\times(\hat{{\Omega}}) F_{b}^\times(\hat{{\Omega}})\right].
\end{equation}

\subsection{Data Sets}\label{subsec:NG_sim}

\subsubsection{NANOGrav 15yr-like data set}\label{subsec:NG_sim}
We base one of our simulated data sets on the NANOGrav 15-year data set, adopting the sky position, timing model, observation times and noise properties for each pulsar \cite{NANOGrav:2023hde}.
From this information, we create \texttt{pint}-based pulsar objects with the epoch-averaged TOAs and uncertainties using \texttt{pta-replicator} and shift the TOAs to perfectly match the timing model template, zeroing out the residuals. Then, we inject white noise, intrinsic red noise, a GWB and a potential hotspot signal into our mock PTA.
We include the intrinsic red noise process for all pulsars that show significant intrinsic red noise even in the presence of a common red noise process in the NANOGrav 15-year data set with amplitudes and spectral indices as given in Table 2 in Ref.~\cite{ng+23_noise}.

After simulating the residuals, the timing model is fit using \texttt{pint} for all pulsars in the PTA. Effectively, this removes the constant and linear components of the timing residuals which mainly originate from noise processes at frequencies $f<1/T_{\rm{obs}}$. Then, we compute the optimal statistic with \texttt{defiant}. For the broadband OS, we follow the choice in the analysis of the NANOGrav 15-year data set and include contributions up to $f_{\rm max}=12/T_{\rm obs}$ \cite{NANOGrav:2023gor}. For the frequency-resolved approach we reconstruct the cross-correlation coefficients in the lowest five frequency bins.

\subsubsection{IPTA DR3-like data set}\label{subsec:IPTA_sim}
To produce a data set with the noise properties of the upcoming IPTA DR3, we consider all 120 pulsars that were part or the most recent data sets of NANOGrav, EPTA, for which we base our simulations on the DR2new subset, MPTA and PPTA. For each pulsar, we adopt the timing model and red noise properties from one of the PTAs, going in the order (1) NANOGrav, (2) EPTA~\cite{EPTA:2023sfo}, with red noise values taken from Table 4 in Ref.~\cite{EPTA:2023akd}, (3) MPTA, where we adopt the noise values from Table 1 of Ref.~\cite{Miles:2024rjc} and (4) PPTA, 
where the red noise parameters are taken from Table 1 in Ref.~\cite{Reardon:2023zen}. We then compute the epoch averaged TOAs and their associated uncertainties for all observation epochs across the individual PTAs timing this pulsar. Finally, we combine all observations into a single data set for this pulsar.

\noindent As the number of cross-correlations increases to $7140$ in this simulated data set, computing the null distributions with the same approach we applied for the NANOGrav-like simulation becomes unfeasible. Instead, we opt for a simpler solution: We do not fit the timing model, as the optimal statistic already marginalizes over linear deviations from the perfect model fit (see, i.e., Sec.~\ref{sec:timing_corr}, Ref.~\cite{NANOGrav:2023icp}), a regime we are guaranteed to be in for simulated pulsars. This removes the requirement for \texttt{pint} based pulsar objects. Instead, we generate idealized, zero-residual \texttt{enterprise} objects and adjust the TOAs and timing residuals directly in those. As the timing model parameters now remain constant, the only change between different realizations of our data sets is given by the shift in TOAs. The effect this shift has on the timing model design matrices and with these the optimal statistic pair-covariance matrices is negligible. They can therefore be assumed to be constant for all realizations and can be precomputed. While this is not a perfect approach for individual samples, as the resulting pair correlations will differ slightly between the exact approach presented in sec. \ref{subsec:NG_sim} and this simplified version, we have verified on NANOGrav-like simulations that the resulting distributions for all quantities calculated with the optimal statistic and all detection statistics are identical.

As for the NANOGrav case, we reconstruct the broadband correlations using twelve frequency bins and consider the first five frequency bins in the narrowband case.

\subsubsection{Idealized PTA}\label{subsec:ideal_sim}
To derive fundamental limits of PTA anisotropy searches, we also generate idealized data sets. 

\noindent The pulsar positions are drawn from uniform distributions. We assume a 14 day observation cadence, equal baselines and same time observations for all pulsars. To achieve results independent of specific timing model, we replace the pulsars' timing model design matrices by a ones where the row corresponding to the $i$-th TOA, $t_{i}$ is given $M_i=(1,t_i,t_i^2)$, effectively projecting out constant, linear and quadratic trends in the timing residuals. The uncertainties on the correlations induced by this choice of timing model design matrix are comparable to the ones appearing with a specific timing model.

\noindent Specifically, we consider a configuration with 100 pulsars and observation baselines of $T_{\rm obs}=15,20,\ldots,50$\,yrs as well as a scenario with a fixed $T_{\rm obs}=30\,\rm{yrs}$ but increasing pulsar numbers $N_{\rm psr}=50,75,\ldots,200$.
The only contribution to the timing residuals in these data sets is given by the gravitational wave background.
As for the IPTA DR3-like data sets, we do not fit for the timing model and assume the pair-covariance matrices to be unchanged under the shift in the TOAs induced by the timing residuals.

We include frequencies from $f_{\min}=1/T_{\rm obs}$ up to a constant cutoff of $f_{\rm max}=1\,\rm{yr}^{-1}$ in the broadband cross-correlation estimation for all variations of these data sets.

\begin{figure}[h!]
\centering
    \textbf{Radiometer basis, $f=2/T_{\rm obs}$}
    \includegraphics{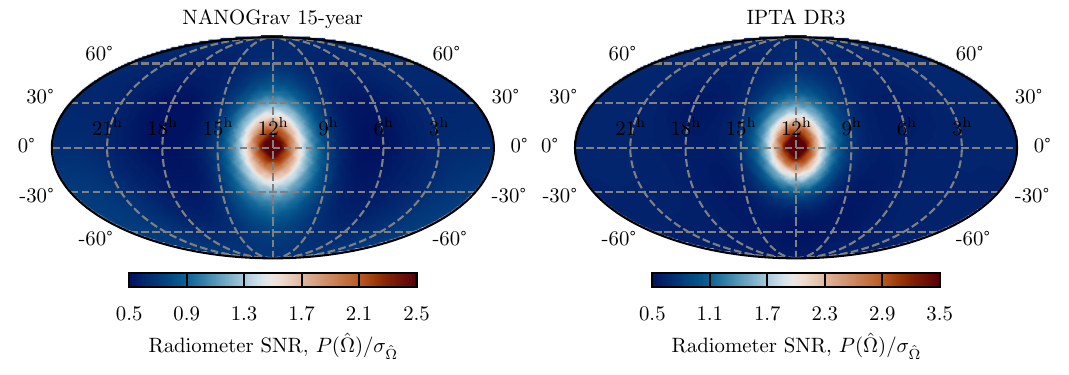}\vspace{2em}
    \textbf{Pixel basis, $f=2/T_{\rm obs}$}
    \includegraphics{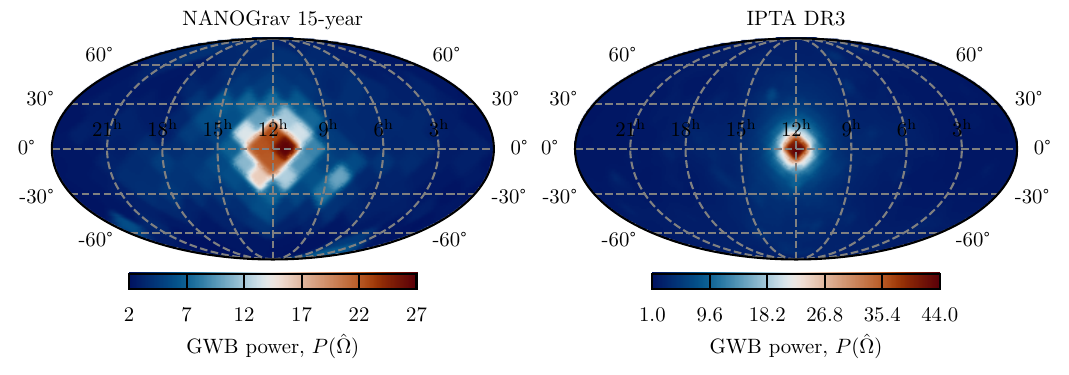}\vspace{2em}
    \textbf{Square-root spherical harmonics basis, $f=2/T_{\rm obs}$}
    \includegraphics{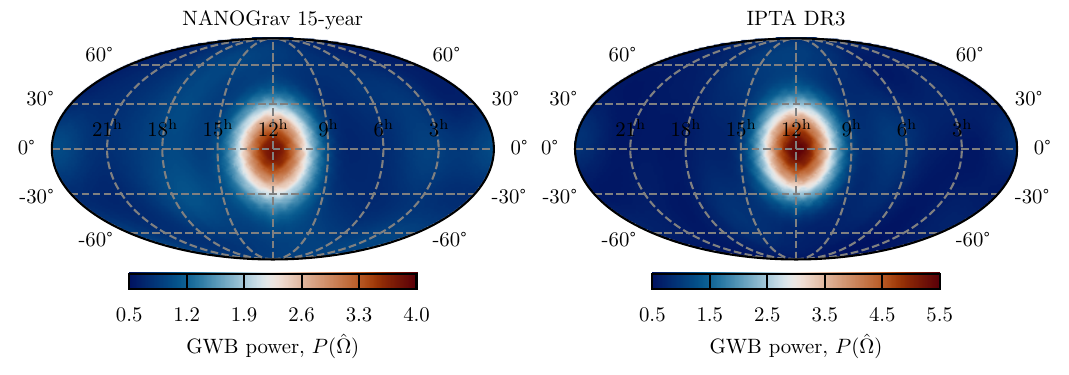}
    \caption{Same as Fig.~\ref{fig:hot_map} but for the frequency-resolved search.}
    \label{fig:hot_map_pf}
\end{figure}

\begin{figure}[t!]
\centering
    \textbf{Radiometer basis, $f=2/T_{\rm obs}$}
    \includegraphics{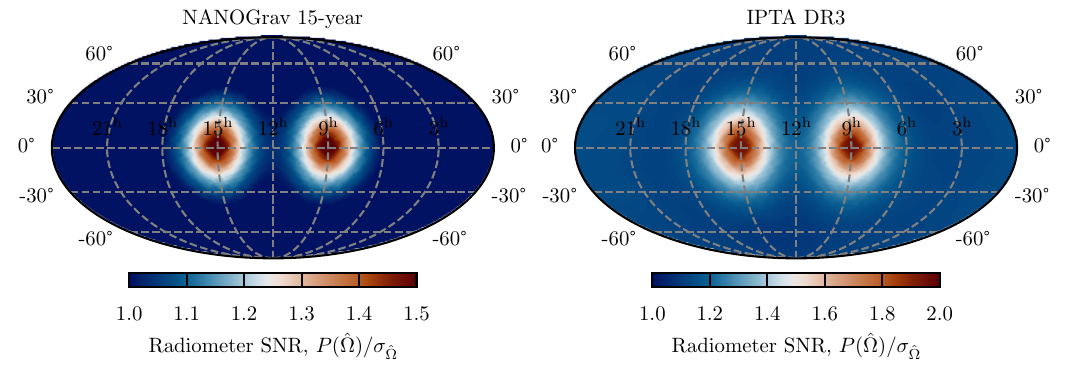}\vspace{2em}
    \textbf{Pixel basis, $f=2/T_{\rm obs}$}
    \includegraphics{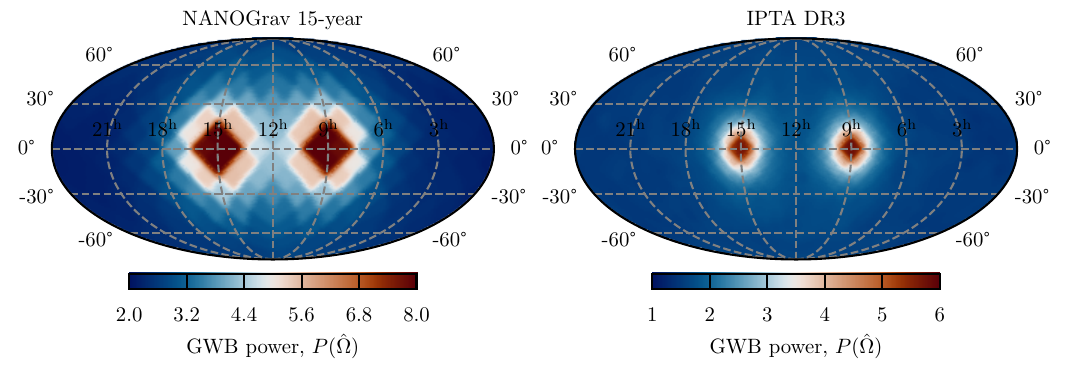}\vspace{2em}
    \textbf{Square-root spherical harmonics basis, $f=2/T_{\rm obs}$}
    \includegraphics{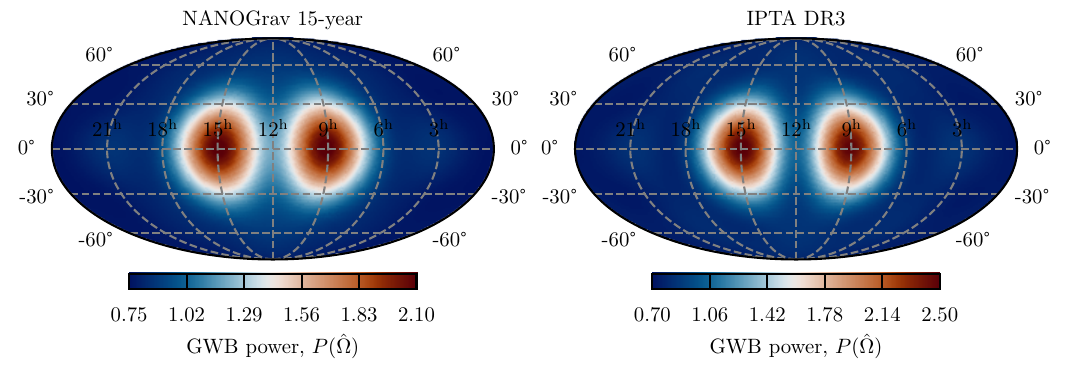}\vspace{2em}
    \caption{Same as Fig.~\ref{fig:hot_map} but for the frequency-resolved search in the case with two hotspots.}
    \label{fig:2hot_map_pf}
\end{figure}

\begin{figure}[h!]
\centering
    \textbf{Radiometer basis, $f=2/T_{\rm obs}$}
    \includegraphics{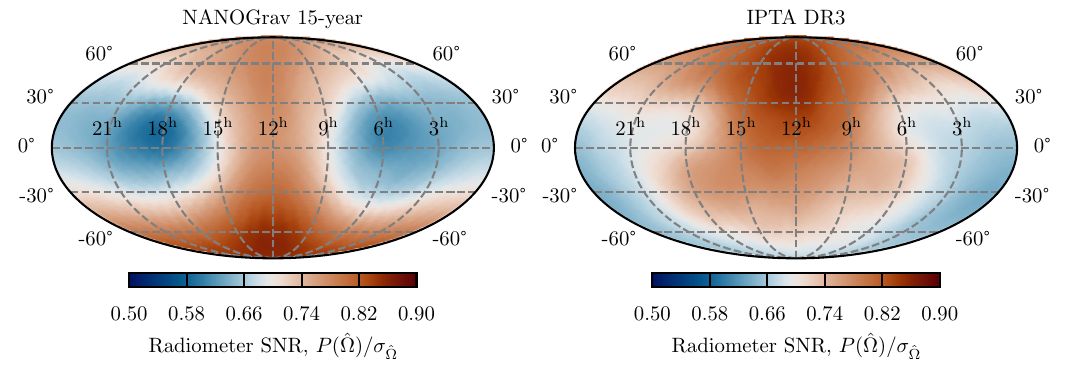}\vspace{2em}
    \textbf{Pixel basis, $f=2/T_{\rm obs}$}
    \includegraphics{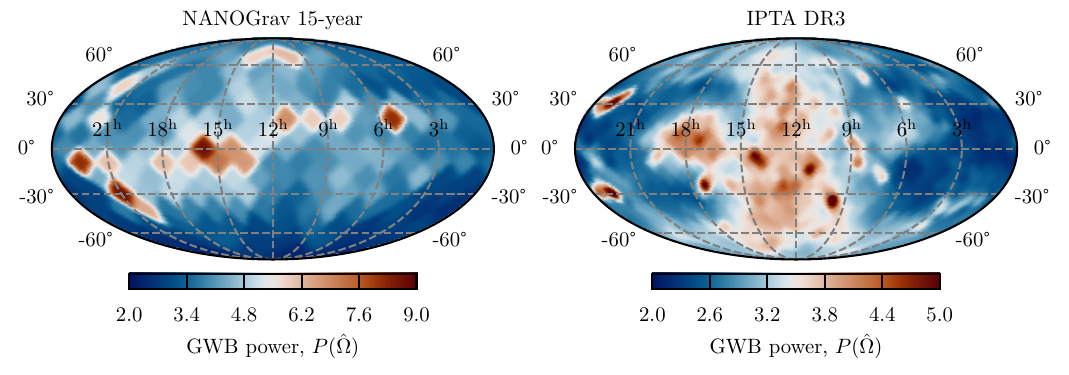}\vspace{2em}
    \textbf{Square-root spherical harmonics basis, $f=2/T_{\rm obs}$}
    \includegraphics{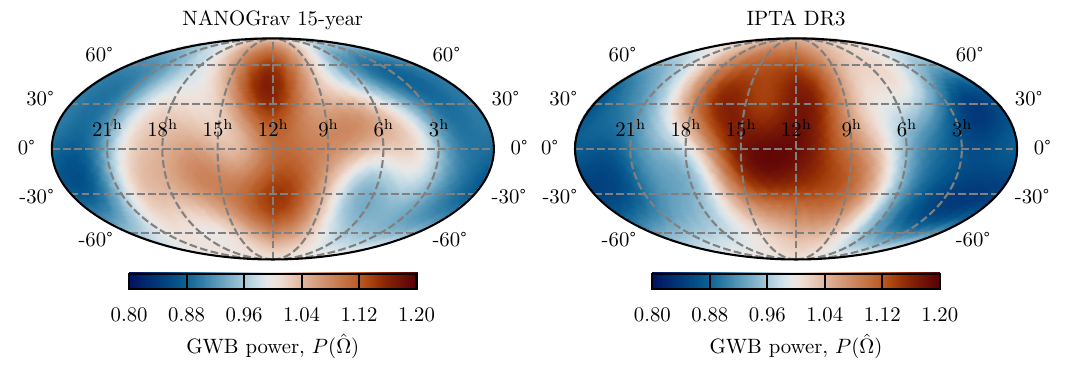}
    \caption{Same as Fig.~\ref{fig:dip_map} but for the frequency-resolved search.}
    \label{fig:dip_map_pf}
\end{figure}

\bibliographystyle{apsrev4-1}
\bibliography{references}
\end{document}